\documentclass[12pt, draftclsnofoot, onecolumn]{IEEEtran}

\usepackage{graphicx}
\usepackage{amsmath,bbm,epsfig,amssymb,amsfonts,  amstext, verbatim,amsopn,cite,subfigure,multirow,multicol,lipsum}
\usepackage{balance}
\usepackage{url}
\usepackage{amsfonts}
\usepackage{epsfig}
\usepackage{setspace}
\usepackage{stmaryrd}
\usepackage{psfrag}	
\usepackage{multirow}
\usepackage{float}
\usepackage[process=auto]{pstool}
\usepackage{etoolbox}
\usepackage{algorithm}
\usepackage{algorithmic}
\usepackage{hyperref}

\usepackage{pifont}
\allowdisplaybreaks

\newtheorem{theo}{Theorem}
\newtheorem{lem}{Lemma}
\newtheorem{remk}{Remark}

\newtheorem{corol}{Corollary}

\newtoggle{OneColumn}

\toggletrue{OneColumn}

\iftoggle{OneColumn}{%
  
}{%
}

\EndPreamble
\begin{document}

\title{Non-Coherent Detection for \\ Diffusive Molecular Communications}

\author{Vahid Jamali\dag, Nariman Farsad\ddag, Robert Schober\dag, and Andrea Goldsmith\ddag\\
\IEEEauthorblockA{\dag Friedrich-Alexander University (FAU), Erlangen, Germany  \\
\ddag Stanford University, Stanford, California, USA \vspace{-1.5cm}
\thanks{This paper has been presented in part at ACM NanoCOM 2016 \cite{NanoCOM16}.} 
\thanks{This work was supported in part by the German Science Foundations (Project SCHO 831/7-1) and the Friedrich-Alexander-
University Erlangen-Nuremberg under the Emerging Fields Initiative (EFI).} 
}
}

\maketitle

\begin{abstract}
We study non-coherent detection schemes for molecular communication (MC) systems that do not require knowledge of the channel state information (CSI). In particular, we first derive the optimal maximum likelihood (ML) multiple-symbol (MS)  detector for MC systems. As a special case of the optimal MS detector, we show that the optimal ML symbol-by-symbol (SS) detector can be equivalently written in the form of a threshold-based detector, where the optimal decision threshold is constant and depends only on  the statistics of the MC channel. The main challenge of the MS detector is the complexity associated with the calculation of the optimal detection metric. To overcome this issue, we propose an approximate MS detection metric which can be expressed in closed form. To reduce complexity even further, we develop a non-coherent decision-feedback (DF) detector and a suboptimal blind detector. Finally, we derive analytical expressions for the bit error rate (BER) of the optimal SS detector, as well as upper and lower bounds for the BER of the optimal MS detector.
Simulation results confirm the analysis and reveal the effectiveness of the proposed optimal and suboptimal detection schemes compared to a benchmark scheme that assumes perfect CSI knowledge, particularly when the number of observations used for detection is sufficiently~large. 
\end{abstract}

\begin{IEEEkeywords} 
Molecular communications, channel state information, non-coherent detection, blind detection, and Poisson channel.
\end{IEEEkeywords} 

\section{Introduction}

Molecular communication (MC) has recently  emerged as a bio-inspired approach for synthetic communication systems having nano/micrometer scale dimensions \cite{Nariman_Survey,Survey_Mol_Nono,InternetMC}. Unlike conventional wireless communication systems that employ electromagnetic waves  to convey information, MC systems encode information in the number,  type, or time of release of signalling molecules. Calcium signaling
of neurons  and the exchange of autoinducers  by bacteria in quorum sensing are among the many examples of MC in nature \cite{Survey_Mol_Nono,CellBio}.

The effect of the diffusive MC channel on a concentration-based MC system is reflected in \textit{i)} the \textit{probability} that a molecule released by the transmitter is observed at the receiver, and \textit{ii)} the \textit{expected} number  of interfering molecules observed  at  the receiver. Hence, \textit{i)} and \textit{ii)} constitute the channel state information (CSI) of the MC channel. The CSI of an MC system depends on parameters such as the diffusion coefficient of the information molecules, the velocity of the flow in the MC channel, the concentration of enzyme in the environment, the distance between the transmitter and the receiver, and the type of the adopted receiver, among other factors, see \cite[Chapters~3 and 4]{Berg} and \cite[Chapter~4]{BioPhysic}.   Most existing works on MC assume that the CSI is  perfectly known at the receiver for reliable detection of the transmitted information bits  \cite{Hamid_Lett,Cl_MF_IEEE,Adam_OptReciever,ConsCIR}. In reality, the CSI is not known a priori and has to be estimated. To this end, a training-based CSI estimation framework was developed in \cite{TCOM_MC_CSI} where several optimal and suboptimal estimators were proposed. However, we note that the CSI of MC systems may change over time due to factors such as variations in the velocity of the flow, the temperature (which leads to variations in the diffusion coefficient and the enzyme concentration), and the distance between the transmitter and the receiver (e.g. if they are suspended in a fluid) \cite{BioPhysic,ArmanMobileMC,ArmanMCStat}. Therefore, the CSI acquisition has to be  conducted repeatedly to keep track of CSI variations.  Detection schemes requiring the acquisition of the CSI are suitable options only if the coherence time of the MC channel is sufficiently large (e.g. when the temperature, the flow, and the positions of the transmitter and the receiver change very slowly) such that the corresponding training overhead is tolerable. On the other hand, for the case when the MC channel changes rapidly,  CSI estimation either entails a large overhead or results in a low CSI estimation quality. In this case, directly detecting the data symbols without spending any resources on CSI acquisition is an attractive option which is referred to as \textit{non-coherent}~detection.

In this paper, our focus is the design of optimal and suboptimal non-coherent  detection schemes that do not require knowledge of the instantaneous CSI. In particular, we first derive the optimal maximum likelihood (ML) multiple-symbol (MS) detector. As a special case of the optimal MS detector, we show that the optimal ML symbol-by-symbol (SS) detector can be equivalently written in the form of a threshold-based detector where the optimal decision threshold is constant and depends only on the statistics of the MC channel. One of the main challenges in implementing the MS detector is the computation of the detection metrics, which requires the evaluation of expectations with respect to the probability density function (PDF) of the CSI. The PDF of the CSI depends on the considered MC environment  and a general analytical expression is not known.  In practice,  the PDF of the CSI for a particular MC channel can be estimated using  empirical measurements of the CSI. To obtain a closed-form expression for the MS detection metric, we approximate the PDF of the CSI by the Gamma distribution. We can show that the Gamma distribution accurately matches the exact distribution of the CSI for several examples of stochastic MC channels. Additionally, to reduce  complexity even further, we develop a non-coherent decision-feedback (DF) detector and a suboptimal blind detector. Furthermore, we derive analytical expressions for the bit error rate (BER) of the optimal SS detector, and an upper bound and a lower bound for the BER of the optimal MS detector.  Our simulation results  reveal the effectiveness of the proposed optimal and suboptimal non-coherent detectors  and show that  their performance approaches that of a benchmark scheme that assumes perfect CSI as the number of symbols used for detection increases. 

In contrast to MC, for conventional wireless communication, there is a rich literature on non-coherent MS detection, see e.g. \cite{NonCoherent_Divsalar,Robert_MSD}, and the references therein. In particular, in \cite{NonCoherent_Divsalar}, MS differential detection without CSI was presented  for radio frequency (RF) communications, and in \cite{Robert_MSD}, non-coherent MS detection for a photon-counting receiver was studied for optical communications. We note that the detection problem and the resulting  detection strategies developed for conventional wireless communications are not applicable to the corresponding MC detection problem due to the fundamental differences in the two systems. The problem of non-coherent data detection in MC was considered before in \cite{NonCoherent_MC,NonCoherent_MC_Convexity} and  heuristic low-complexity non-coherent symbol-by-symbol detectors were proposed.  However, to the best of the authors' knowledge, the design of  \textit{non-coherent MS detectors},  which is considered in this paper and its conference version \cite{NanoCOM16}, has not yet been considered in the MC literature.  In addition to the detectors proposed in \cite{NanoCOM16}, this paper derives the BER of the optimal SS detector along with lower and upper bounds for the BER of the optimal MS detector, and validates these results with simulations.
 
The  remainder  of  this  paper  is  organized  as  follows.  
In Section~II, the  system model and the CSI model  are  introduced.
The proposed detectors are presented in Section~III and their performance is analyzed in Section~IV. Numerical results are presented in Section~V, and conclusions are drawn in Section~VI.

\textit{Notation:} We use the following notation throughout this paper: $\mathbbmss{E}_{x}\{\cdot\}$ denotes expectation with respect to random variable (RV) $x$ and $|\cdot|$ represents the cardinality of a set.   Bold letters are used to denote vectors, $\mathbf{a}^T$ represents the transpose of vector $\mathbf{a}$, and $\mathbf{a}\geq\mathbf{0}$ specifies that all elements of vector $\mathbf{a}$ are non-negative. Moreover, $\mathbb{N}$ and $\mathbb{R}^+$ denote the sets of non-negative integer  numbers and positive real numbers, respectively.
 $\lfloor\cdot\rfloor$ and $\lceil\cdot\rceil$ denote the floor and ceiling functions which map a real number to the largest previous and the smallest following integer number,  respectively. $\Gamma(\cdot)$ is the Gamma function and $\mathbf{1}\{\cdot\}\in\{0,1\}$ is  an  indicator  function  which  is  equal  to one  if  the  argument  is  true  and  equal  to  zero  if  it  is  not
true. Moreover,  $\mathrm{Poiss}(\lambda)$  denotes a Poisson RV with mean  $\lambda$, $\mathrm{Bin}(n,p)$ denotes a binomial RV for $n$  trials and success probability $p$, $\mathcal{N}\left(\mu,\sigma^2\right)$ denotes a Gaussian RV with mean $\mu$ and variance $\sigma^2$, and $\mathrm{Gamma}(\alpha,\beta)$ denotes a Gamma distributed RV with scale parameter $\alpha$ and rate parameter~$\beta$.

\section{System Model}

In this section, we introduce the MC channel model and the CSI model used in this paper. 

\subsection{Channel Model}

\begin{figure}
  \centering
 \scalebox{0.6}{
\pstool[width=1\linewidth]{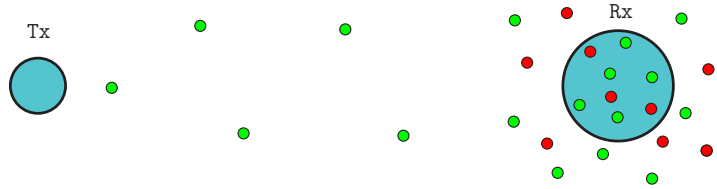}{
\psfrag{T}[c][c][1.2]{$\mathtt{Tx}$}
\psfrag{R}[c][c][1.2]{$\mathtt{Rx}$}
} } \vspace{-0.3cm}
\caption{Schematic illustration of the considered MC system where the  molecules released by the transmitter in a given symbol interval are shown in green  color whereas  the  noise molecules are shown in red color.\vspace{-0.3cm}}
\label{Fig:TxRx}
\end{figure}

We consider an MC system  consisting of a transmitter, a channel, and a receiver, see Fig.~\ref{Fig:TxRx}. At the beginning of each symbol interval, the transmitter releases either $N^{\mathtt{TX}}$ or zero molecules corresponding to the  binary bits $1$ and $0$, respectively, i.e., the modulation format used is ON-OFF keying (OOK) \cite{Nariman_Survey}. In this paper, we assume that the transmitter emits only one type of molecule. The released molecules diffuse through the fluid medium between the transmitter and the receiver.  The receiver counts the number of observed molecules in each symbol interval.  We note that this is a rather general model for the MC receiver which includes well-known receivers such as the transparent receiver \cite{ConsCIR}, the absorbing receiver \cite{Chae_Absorbing}, and the reactive receiver \cite{Arman_ReactReciever}.
In particular, the number of observed molecules at the receiver  in symbol interval $k$, denoted by $r[k]$, is given~by \cite{Yilmaz_Poiss,HamidJSAC,TCOM_MC_CSI,ICC2017_MC_IEEE}
\begin{align} \label{Eq:ChannelInOut}
  r[k]  =  c_{\mathtt{s}}[k] + c_{\mathtt{n}}[k],
\end{align}
where $c_{\mathtt{s}}[k]$ is the number of  molecules observed at the receiver  in symbol interval $k$ due to the release of $s[k]N^{\mathtt{Tx}}$ molecules by the transmitter at the beginning of symbol interval $k$, with $s[k]\in\{0,1\}$. We assume that the binary information bits are equiprobable, i.e., $\Pr\{s[k]=1\}=\Pr\{s[k]=0\}=0.5$.  Moreover, $c_{\mathtt{n}}[k]$ is the number of interfering noise molecules comprising residual inter-symbol interference (ISI), multiuser interference (caused by other MC links), and  external noise (originating from natural sources) observed by the receiver in symbol interval~$k$.

The MC channel is dispersive due to the diffusive propagation of the molecules \cite{Adam_Enzyme}. The ISI-free communication model in (\ref{Eq:ChannelInOut}) implies that  the symbol intervals are chosen sufficiently large such that the channel impulse response (CIR) fully decays to zero within one symbol interval. We note that enzymes  \cite{Adam_Enzyme} and reactive information molecules, such as acid/base molecules \cite{Nariman_Acid}, may be used to speed up the decaying of the CIR as a function of time. Nevertheless, since the length of the symbol intervals is finite, some residual ISI always exists. Throughout this  paper, we assume that the effect of the residual ISI is included in $c_{\mathtt{n}}[k]$ and is sufficiently small compared to the other components of $c_{\mathtt{n}}[k]$ such that $c_{\mathtt{n}}[k]$ is (approximately) independent of the signal $c_{\mathtt{s}}[k] $.  We now describe the underlying models and associated distributions of RVs $c_{\mathtt{s}}[k] $ and $ c_{\mathtt{n}}[k]$.

\textit{Information Molecules ($c_{\mathtt{s}}[k]$):} The movements of individual molecules are assumed to be independent from each other. We further assume that  the observations of different molecules at the receiver are independent\footnote{This assumption holds for the transparent and absorbing receiver models. For this assumption to hold for the reactive receiver model in \cite{Arman_ReactReciever}, the number of receptors on the surface of the receiver has to be much larger than the average number of molecules around the receiver which is a valid assumption for typical biological cells in nature, see \cite[Page~33]{Berg}.}. Therefore, from a probabilistic point of view, we can assume that each molecule released by the transmitter in a given symbol interval is observed at the receiver in the same symbol interval with a certain probability, denoted by $p_{\mathtt{s}}$. Since any given molecule released by the transmitter is either observed by the receiver or not, a binary state model applies and the number of observed molecules follows the binomial distribution $\mathrm{Bin}(N^{\mathtt{Tx}},p_{\mathtt{s}})$. Moreover, assuming that $N^{\mathtt{Tx}}$ is very large while $N^{\mathtt{Tx}}p_{\mathtt{s}}\triangleq \bar{c}_{\mathtt{s}}$ is relatively small, the  binomial distribution $\mathrm{Bin}(N^{\mathtt{Tx}},p_{\mathtt{s}})$ converges to the Poisson distribution $\mathrm{Poiss}\left(\bar{c}_{\mathtt{s}}\right)$ \cite{BayesianBook}. Another approximation of the binomial distribution $\mathrm{Bin}(N^{\mathtt{Tx}},p_{\mathtt{s}})$ is the Gaussian distribution $\mathcal{N}(N^{\mathtt{Tx}}p_{\mathtt{s}},\sqrt{N^{\mathtt{Tx}}p_{\mathtt{s}}(1-p_{\mathtt{s}})})$ which holds for very large $N^{\mathtt{Tx}}$ when $\bar{c}_{\mathtt{s}}$ is relatively large  \cite{BayesianBook}. We note that the assumptions for the Poisson approximation are more justified for MC than those of the Gaussian approximation since the number of released molecules is typically very large (on the order of hundreds or a few thousands) but, typically, only few molecules reach the receiver. Therefore, we adopt the Poisson approximation in this paper, i.e., $c_{\mathtt{s}}[k]\sim\mathrm{Poiss}\left(\bar{c}_{\mathtt{s}} s[k]\right)$. The accuracy of the Poisson distribution in modeling the number of  molecules observed at the receiver was verified in \cite{Yilmaz_Poiss}, \cite{Hamid_Lett,Adam_OptReciever}  and compared with the accuracy of the corresponding binomial and Gaussian models\footnote{For instance,  the analytical framework developed in \cite{Yilmaz_Poiss} can be used to show that for $N^{\mathtt{Tx}}=1000$, if $p_{\mathtt{s}}\leq 0.115$ holds, the Poisson distribution more accurately approximates the binomial distribution  in terms of the root mean squared error (RMSE) of the cumulative distribution function (CDF), whereas, if $p_{\mathtt{s}}>0.115$ holds, the Gaussian approximation is a better fit. For MC systems, if  $N^{\mathtt{Tx}}=1000$ molecules are released by the transmitter, typically we expect to observe much fewer than $N^{\mathtt{Tx}}p_{\mathtt{s}}=113$ molecules at the receiver. Hence, the Poisson approximation is more accurate compared to the Gaussian approximation in~this~case.}. 

\textit{Noise Molecules ($c_{\mathtt{n}}[k]$):}  Noise molecules originate from  interfering natural or synthetic sources  \cite{Adam_Universal_Noise}. Suppose there are in total $N^{\mathtt{noise}}$   noise molecules in the environment and they are uniformly distributed in space and time, which is a reasonable assumption if a priori knowledge about the locations and activities of the noise sources is not available. In addition,  the probability that at any given time, any noise molecule is observed at the receiver is assumed to be the same for all noise molecules and is denoted by $p_{\mathtt{n}}$.  Since at the sampling time, each noise molecule is either observed at the receiver or not, again, a two state model is valid and $c_{\mathtt{n}}[k]$ follows a binomial distribution, i.e., $c_{\mathtt{n}}[k]\sim\mathrm{Bin}(N^{\mathtt{noise}},p_{\mathtt{n}})$. Moreover, the probability that a given noise molecule is observed at the receiver is again expected to be very small, i.e.,  $p_{\mathtt{n}}\to 0$ holds. Therefore,  $N^{\mathtt{noise}}\to \infty$ has to hold such that the average number of noise molecules observed at the receiver, denoted by $\bar{c}_{\mathtt{n}}=N^{\mathtt{noise}} p_{\mathtt{n}}$, is non-zero. Considering this, $c_{\mathtt{n}}[k]$ follows also a Poisson distribution, i.e., $c_{\mathtt{n}}[k]\sim\mathrm{Poiss}\left(\bar{c}_{\mathtt{n}}\right)$.

\begin{remk}
The channel model in (\ref{Eq:ChannelInOut}) can be generalized to the case of multiple-sample detection if the following sum detector is employed:
\begin{align} \label{Eq:WeightedSum}
 r[k] &=  \sum_{m=1}^M  y[k,m] =  \sum_{m=1}^M  c_{\mathtt{s}}[k,m] +  \sum_{m=1}^M  c_{\mathtt{n}}[k,m] 
 \triangleq c_{\mathtt{s}}[k] + c_{\mathtt{n}}[k],
\end{align}
where $M$ denotes the number of samples per symbol interval and $y[k,m]$ is the number of molecules observed at the receiver in the $m$-th sample of symbol interval $k$. Moreover, $c_{\mathtt{s}}[k,m]$ is the number of  molecules observed at the receiver for the $m$-th sample of symbol interval $k$ due to the release of $s[k]N^{\mathtt{Tx}}$ molecules by the transmitter at the beginning of symbol interval $k$,  and $c_{\mathtt{n}}[k,m]$ is the number of noise molecules observed at the receiver for the $m$-th sample of symbol interval~$k$. Since $c_{\mathtt{s}}[k,m]$ and $c_{\mathtt{n}}[k,m]$ are Poisson RVs,  $c_{\mathtt{s}}[k]$ and $c_{\mathtt{n}}[k]$ follow Poisson distributions with means  $\bar{c}_{\mathtt{s}}= \sum_{m=1}^M  \bar{c}_{\mathtt{s}}^{(m)}$ and $M\bar{c}_{\mathtt{n}}$, respectively, where $\bar{c}_{\mathtt{s}}^{(m)}=\mathbbmss{E}\left\{c_{\mathtt{s}}[k,m]\right\}$ and $\bar{c}_{\mathtt{n}}=\mathbbmss{E}\left\{c_{\mathtt{n}}[k,m]\right\}$. We note that the sum detector in (\ref{Eq:WeightedSum}) includes the well-known peak value \cite{Adam_OptReciever} and energy \cite{DistanceEstLett} detectors as special cases when only one sample at the peak concentration is taken and (ideally infinitely) many samples per symbol interval are taken, respectively. 
\end{remk}

\begin{remk}
 Unlike the conventional linear input-output  model for channels in wireless communications \cite{NonCoherent_Divsalar}, the channel model in (\ref{Eq:ChannelInOut}) is not linear since $s[k]$ does not affect the observation $r[k]$ directly but rather via the Poisson RV $c_{\mathtt{s}}[k]$. However, the \textit{expectation} of the received signal is linearly dependent on the transmitted signal, i.e.,
\begin{align} \label{Eq:AveInOut}
 \bar{r}[k] = \mathbbmss{E}\left\{r[k]\right\} =  \bar{c}_{\mathtt{s}} s[k] + \bar{c}_{\mathtt{n}}.
\end{align}
We note that for a given $s[k]$, in general, the actual number of molecules observed at the receiver, $r[k]$, will differ from the expected number of observed molecules, $\bar{r}[k]$, due to the intrinsic noisiness of diffusion. 
\end{remk}

\subsection{CSI Model}

The state of the diffusive MC channel specified in (\ref{Eq:ChannelInOut}) can be captured by $p_{\mathtt{s}}$ and $\bar{c}_{\mathtt{n}}$. Since we employ OOK signaling and $N^{\mathtt{Tx}}$ is assumed to be fixed, without loss of generality, we refer to the vector $\bar{\mathbf{c}}=[\bar{c}_{\mathtt{s}},\bar{c}_{\mathtt{n}}]^T$ (instead of $[p_{\mathtt{s}},\bar{c}_{\mathtt{n}}]^T$) as the CSI of the considered MC system in the remainder of this paper. 
Most existing detection schemes in MC assume that knowledge of the CSI is available at the receiver \cite{Hamid_Lett,Adam_OptReciever,ConsCIR}. In contrast, in this paper,  we directly detect a block of transmitted symbols  based on the corresponding received observations without spending any resources on CSI acquisition at the receiver.  Let $\mathbf{s}=[s[1],s[2],\dots,s[K]]^T$ and $\mathbf{r}=[r[1],r[2],\dots,r[K]]^T$ denote the vectors of the transmitted symbols and the received observations over a block of $K$ symbol intervals, respectively. We assume that the CSI remains unchanged over one block of transmitted symbols, but may change from one block to the next (e.g., due to a change in MC channel parameters such as the temperature, the velocity of the flow, etc.). To model this, we assume that  CSI, $\bar{\mathbf{c}}$, is an RV that takes its values in each block according to PDF  $f_{\bar{\mathbf{c}}}(\bar{\mathbf{c}})$. Furthermore, we assume that RVs $\bar{c}_{\mathtt{s}}$ and $\bar{c}_{\mathtt{n}}$ are independent, i.e., $f_{\bar{\mathbf{c}}}(\bar{\mathbf{c}})=f_{\bar{c}_{\mathtt{s}}}(\bar{c}_{\mathtt{s}}) f_{\bar{c}_{\mathtt{n}}}(\bar{c}_{\mathtt{n}})$ where $f_{\bar{c}_{\mathtt{s}}}(\bar{c}_{\mathtt{s}}) $ and $ f_{\bar{c}_{\mathtt{n}}}(\bar{c}_{\mathtt{n}})$ are the marginal PDFs of $\bar{c}_{\mathtt{s}}$ and $\bar{c}_{\mathtt{n}}$, respectively.  For future reference, in the rest of this work,  we define $ f_{\mathbf{r}}(\mathbf{r}|\bar{\mathbf{c}},\mathbf{s}) = \prod_{k} f_{r[k]}(r[k]|\bar{\mathbf{c}},s[k])$ as the PDF of the observation  vector $\mathbf{r}$  conditioned on both CSI $\bar{\mathbf{c}}$ and  transmitted symbol vector $\mathbf{s}$,  and we define $f_{\mathbf{r}}(\mathbf{r}|\mathbf{s}) = \prod_k f_{r[k]}(r[k]|s[k])$ as the PDF of  $\mathbf{r}$ conditioned  only on $\mathbf{s}$. We note that although the proposed non-coherent detection schemes do not require knowledge of the instantaneous CSI, $\bar{\mathbf{c}}$, we assume that statistical CSI, i.e., $f_{\bar{\mathbf{c}}}(\bar{\mathbf{c}})$,  is available for the design of the proposed non-coherent MS, SS, and DF detectors.  Since obtaining the CSI statistics might be difficult for some practical systems, we  also propose a suboptimal blind detector that does not require  statistical CSI knowledge.

\begin{remk}
We note that for the conventional channel model used in wireless communications, the noise power depends  largely on the characteristics of the receiver  \cite{NonCoherent_Divsalar,Robert_MSD}. For instance, in RF communications, the noise variance is determined by the receiver thermal noise \cite{NonCoherent_Divsalar}, and in optical communications, the power of the noise depends on the sensitivity of the photo-detector to the background radiation \cite{Robert_MSD}. Therefore, when modeling these systems, it is commonly assumed that the noise power is constant. Hence, the channel gain  is typically referred to as the CSI and the noise power is assumed to be fixed and known. On the other hand, the noise molecules in MC originate from  interfering natural or synthetic sources. Thus, the mean of the noise is in general dependent on the properties of the MC channel in addition to the type of the adopted receiver.  Hence, it is reasonable to treat both $\bar{c}_{\mathtt{s}}$ and $\bar{c}_{\mathtt{n}}$  as the CSI of the MC~system. 
\end{remk}

\section{Optimal and Suboptimal Non-Coherent Detection}
In this section, we first present the optimal coherent detector which provides a performance upper bound for any non-coherent detector.   Subsequently, we present the proposed non-coherent detectors, namely the optimal non-coherent MS, SS, and DF detectors. Furthermore, motivated by the structure of the optimal coherent detector, we develop a simple blind detector.

\subsection{Coherent ML Detector}

As a performance upper bound, we consider the optimal detector for perfect CSI knowledge. Moreover, since the observations in different symbol intervals are independent, without loss of optimality, the considered benchmark scheme performs symbol-by-symbol detection. Thus, the optimal ML detector is given by
\begin{align} \label{Eq:ML_CIR}
  \hat{s}^{\mathtt{ML}}[k] &= \underset{s[k]\in\{0,1\}}{\mathrm{argmax}} \,\,f_{r[k]}\big(r[k] \big |\bar{\mathbf{c}},s[k]\big) \nonumber \\
  &= \underset{s[k]\in\{0,1\}}{\mathrm{argmax}} \,\,\frac{\left( \bar{c}_{\mathtt{s}} s[k] + \bar{c}_{\mathtt{n}} \right)^{r[k]} \exp\left(- \bar{c}_{\mathtt{s}} s[k] - \bar{c}_{\mathtt{n}} \right)}{r[k]!},
\end{align}
where $f_{r[k]}\big(r[k] \big |\bar{\mathbf{c}},s[k]\big)$ is the Poisson distribution function. The ML detector can be rewritten in the form of  a threshold-based detector as follows \cite{HamidJSAC}
\begin{align} \label{Eq:ML_CIR_Thr}
  \hat{s}^{\mathtt{ML}}[k] = \begin{cases}
  1,\quad &\mathrm{if}\,\, r[k] \geq \xi^{\mathtt{ML}}(\bar{\mathbf{c}}) \\
  0,\quad &\mathrm{otherwise} \\
  \end{cases}
\end{align}
where the computation of the decision threshold, $\xi^{\mathtt{ML}}(\bar{\mathbf{c}}) = \frac{\bar{c}_{\mathtt{s}}}{\mathrm{ln}(1+\frac{\bar{c}_{\mathtt{s}}}{\bar{c}_{\mathtt{n}}})}$, requires CSI knowledge.

\subsection{Non-Coherent ML Detector}
We first develop the optimal non-coherent MS and SS detectors. Following this development, we propose suboptimal decision metrics to cope with the complexity of evaluating the optimal MS and SS metrics for this detector.

\subsubsection{Optimal MS Detector}
We first formulate the ML problem for non-coherent detection of a block of symbols. Note that, in this case, joint detection of  multiple symbols  is beneficial, despite the fact that the symbols are independent, since observing multiple symbols provides more information about the MC channel than observing just one symbol. The optimal ML MS detector is mathematically given~by
\begin{align} \label{Eq:MLMS}
  \hat{\mathbf{s}}^{\mathtt{MS}} &= \underset{\mathbf{s}\in\mathcal{A}}{\mathrm{argmax}} \,\,f_{\mathbf{r}}(\mathbf{r}|\mathbf{s}) 
  = \underset{\mathbf{s}\in\mathcal{A}}{\mathrm{argmax}}  \int_{\bar{\mathbf{c}}\geq \mathbf{0}} f_{\mathbf{r}}(\mathbf{r}|\bar{\mathbf{c}},\mathbf{s})f_{\bar{\mathbf{c}}}(\bar{\mathbf{c}}) \mathrm{d}\bar{\mathbf{c}} \nonumber \\
& = \underset{\mathbf{s}\in\mathcal{A}}{\mathrm{argmax}}   \int_{\bar{c}_{\mathtt{s}}\geq 0} \int_{\bar{c}_{\mathtt{n}}\geq 0} \prod_{k=1}^{K} \frac{\left( \bar{c}_{\mathtt{s}} s[k] + \bar{c}_{\mathtt{n}} \right)^{r[k]} \exp\left(- \bar{c}_{\mathtt{s}} s[k] - \bar{c}_{\mathtt{n}} \right)}{r[k]!} \times f_{\bar{c}_{\mathtt{s}}}(\bar{c}_{\mathtt{s}})  f_{\bar{c}_{\mathtt{n}}}(\bar{c}_{\mathtt{n}}) \mathrm{d}\bar{c}_{\mathtt{s}}\mathrm{d}\bar{c}_{\mathtt{n}}, 
\end{align}
where $\mathcal{A}$ is the set of all $2^K$ possible  binary sequences of length $K$. In (\ref{Eq:MLMS}), we employ the multivariate Poisson distribution function $f_{\mathbf{r}}(\mathbf{r}|\bar{\mathbf{c}},\mathbf{s})$ and  exploit the fact that the observations in different symbol intervals are independent and that RVs $\bar{c}_{\mathtt{s}}$ and $\bar{c}_{\mathtt{n}}$ are independent. Before presenting the optimal MS detector as a solution of (\ref{Eq:MLMS}) in the following theorem, we introduce some auxiliary variables. For a given hypothetical sequence $\mathbf{s}$, let  $\mathcal{K}_1$ and $\mathcal{K}_0$ denote the sets of indices $k$ for which $s[k]=1$ and $s[k]=0$ holds, respectively. Moreover, we define $\mathbf{n}=[n_1,n_0]^T$ where $n_1$ and $n_0$ are the number of ones and zeros in the given hypothetical sequence $\mathbf{s}$. Additionally, for a given observation vector $\mathbf{r}$, we define  $\mathbf{N}=[N_1,N_0]^T$ where $N_1=\sum_{k\in\mathcal{K}_1} r[k]$ and $N_0=\sum_{k\in\mathcal{K}_0} r[k]$.

\begin{theo}\label{Theo:MLMS}
The optimal non-coherent MS detector as the solution to (\ref{Eq:MLMS}) selects a sequence whose ``1" elements  correspond to the $\zeta^{\mathtt{MS}}(\mathbf{r})$ largest elements of $\mathbf{r}$. Moreover, the optimal threshold  $\zeta^{\mathtt{MS}}(\mathbf{r})$ is obtained as
\begin{align} \label{Eq:MLMS_Sol}
\zeta^{\mathtt{MS}}(\mathbf{r}) = \underset{ n_1 \in\{0,1,\dots,K\} }{\mathrm{argmax}} \,\, \Lambda^{\mathtt{MS}}(\mathbf{s},\mathbf{r}),
\end{align}
where $\Lambda^{\mathtt{MS}}(\mathbf{s},\mathbf{r})$ is the MS detection metric given by
\begin{align} \label{Eq:MLMS_Metric}
\Lambda^{\mathtt{MS}}(\mathbf{s},\mathbf{r})  &= \mathbbmss{E}_{\bar{\mathbf{c}}}\Big\{ 
  \left( \bar{c}_{\mathtt{s}} + \bar{c}_{\mathtt{n}} \right)^{N_1}
  \bar{c}_{\mathtt{n}}^{N_0}\, e^{-n_1\bar{c}_{\mathtt{s}}  - K \bar{c}_{\mathtt{n}} } \Big\} \nonumber \\
  &=\sum_{i=0}^{N_1} {N_1 \choose i} \mathbbmss{E}_{\bar{c}_{\mathtt{s}}}\left\{
   \bar{c}_{\mathtt{s}}^{N_1-i} e^{- n_1 \bar{c}_{\mathtt{s}}} \right\} 
    \mathbbmss{E}_{\bar{c}_{\mathtt{n}}}\left\{   \bar{c}_{\mathtt{n}}^{N_0+i} e^{- K \bar{c}_{\mathtt{n}}}   \right\}.\quad
\end{align}
Furthermore, as $K\to\infty$, we obtain  $\zeta^{\mathtt{MS}}(\mathbf{r})\to K\times\Pr\{s[k]=1\}=\frac{K}{2}$.
\end{theo}

\begin{IEEEproof}
The proof is provided in Appendix~\ref{App:Theo_MLMS}.
\end{IEEEproof}


\begin{remk}
The complexity of the proposed  MS detector in Theorem~\ref{Theo:MLMS} is significantly smaller than that of the full search required in (\ref{Eq:MLMS}). In particular, the complexity of the full search in (\ref{Eq:MLMS}) grows exponentially in $K$, i.e., $|\mathcal{A}|=2^K$, whereas the complexity of the search in Theorem~\ref{Theo:MLMS} is linearly increasing in $K$, i.e., there are $K+1$ possibilities.  Furthermore, for each search step, we have to calculate the metric $\Lambda^{\mathtt{MS}}(\mathbf{s},\mathbf{r})$ which is a function of the statistical CSI, but not of the instantaneous CSI, since  $\bar{c}_{\mathtt{s}}$ and $\bar{c}_{\mathtt{n}}$ are averaged out in $\Lambda^{\mathtt{MS}}(\mathbf{s},\mathbf{r})$, cf. (\ref{Eq:MLMS_Metric}).  Moreover, all expectations in the MS detection metric are in the form of $\mathbbmss{E}_{x}\left\{x^{a}e^{-bx}\right\}$ where $x\in\{\bar{c}_{\mathtt{s}},\bar{c}_{\mathtt{n}}\}$ and $a$ and $b$ are constants.  Therefore, if these expectations for all required $a$ and $b$ can be computed offline, they can be stored at the receiver and used for online data detection.  
\end{remk}

As a special case of the optimal MS detector in Theorem~\ref{Theo:MLMS}, in the following corollary we present the optimal SS detector which, unlike the general form in Theorem~\ref{Theo:MLMS}, lends itself to a simple threshold-based detection.

\begin{corol}\label{Corol:MLSS}
The optimal non-coherent SS detector as the solution to (\ref{Eq:MLMS}) for $K=1$ can be written in the form of the following threshold-based detector: 
\begin{IEEEeqnarray}{lll} \label{Eq:MLSS_sol}
  \hat{s}^{\mathtt{SS}}[k] = \begin{cases}
  1,\quad &\mathrm{if}\,\, r[k] \geq \xi^{\mathtt{SS}} \\
  0,\quad &\mathrm{otherwise} \\
  \end{cases}
\end{IEEEeqnarray}
where threshold $\xi^{\mathtt{SS}}$  is given by
\begin{IEEEeqnarray}{lll} \label{Eq:MLSS_thr}
 \xi^{\mathtt{SS}} &= \left\lceil \xi\in\mathbb{R}^+ \big| \mathbbmss{E}_{\bar{\mathbf{c}}}\big\{ 
  \left( \bar{c}_{\mathtt{s}} + \bar{c}_{\mathtt{n}} \right)^{\xi}
    e^{-\bar{c}_{\mathtt{s}}  - \bar{c}_{\mathtt{n}}} \big\} = \mathbbmss{E}_{\bar{c}_{\mathtt{n}}}\big\{ 
   \bar{c}_{\mathtt{n}}^{\xi}
    e^{ - \bar{c}_{\mathtt{n}}} \big\}\right\rceil \nonumber \\
&=\min \left\{ \xi \in\mathbb{N} \Big| \sum\limits_{i = 0}^{\xi}  {\xi \choose i} 
  \mathbbmss{E}_{\bar{c}_{\mathtt{s}}}\left\{ \bar{c}_{\mathtt{s}}^{\xi-i}e^{-\bar{c}_{\mathtt{s}}} \right\}
   \mathbbmss{E}_{\bar{c}_{\mathtt{n}}}\left\{ \bar{c}_{\mathtt{n}}^{i}e^{-\bar{c}_{\mathtt{n}}} \right\} 
   > \mathbbmss{E}_{\bar{c}_{\mathtt{n}}}\left\{ \bar{c}_{\mathtt{n}}^{\xi} e^{ - \bar{c}_{\mathtt{n}} } \right\} \right\}.
\end{IEEEeqnarray}
\end{corol}

\begin{IEEEproof}
The proof is provided in Appendix~\ref{App:Corol_MLSS}.
\end{IEEEproof}

Corollary~\ref{Corol:MLSS} reveals that, similar to the coherent ML detector under perfect CSI knowledge in (\ref{Eq:ML_CIR_Thr}), the optimal non-coherent SS detector has a threshold-based structure. However, the threshold for the optimal SS detector depends only on the statistics of the MC channel. Therefore, $\xi^{\mathtt{SS}}$ remains fixed as long as the statistics of the MC channel do not change. Hence, $\xi^{\mathtt{SS}}$ can be obtained offline once and thereafter used for online detection. Consequently, the optimal SS detector in Corollary~\ref{Corol:MLSS} is considerably less complex than the equivalent detector obtained from Theorem~\ref{Theo:MLMS} for $K=1$, which requires the online computation of threshold $\zeta^{\mathtt{MS}}(r[k])$  or, equivalently, detection metrics $\Lambda^{\mathtt{MS}}(s[k]=1,r[k])$ and $\Lambda^{\mathtt{MS}}(s[k]=0,r[k])$ in each symbol~interval.

\subsubsection{Suboptimal ML Metric}

The main challenge of the optimal MS detector  is the  calculation of the detection metric in (\ref{Eq:MLMS_Metric}). In particular, the detection metric in (\ref{Eq:MLMS_Metric}) involves expectations which require the PDF of the CSI, i.e., $f_{x}(x),\,\,x\in\{\bar{c}_{\mathtt{s}},\bar{c}_{\mathtt{n}}\}$. These PDFs  depend on the considered MC environment  and  general analytical expressions for $f_{x}(x),\,\,x\in\{\bar{c}_{\mathtt{s}},\bar{c}_{\mathtt{n}}\}$, are not known.  In practice,   for a particular MC channel, these PDFs can be obtained using  empirical measurements of $\bar{c}_{\mathtt{s}}$ and $\bar{c}_{\mathtt{n}}$. However, the empirical PDFs might not lend themselves to a simple analytical form. Therefore, one convenient approach for obtaining a mathematical expression for the detection metric in  (\ref{Eq:MLMS_Metric}) is to assume a particular parametric model for $f_{x}(x)$ and to adjust the parameters of the model to match the simulation/experimental data. In general, it is desirable that a parametric PDF model, $f_x^{\mathtt{para}}(x)$, meets the following criteria:
\begin{itemize}
\item $f_x^{\mathtt{para}}(x)$ should meet any inherent constraints of the RV $x$. Such constraints might include that $x$ is non-negative or non-positive, or that it is discrete, continuous, or mixed. 
\item $f_x^{\mathtt{para}}(x)$ should facilitate analysis and ideally lead to closed-form expressions for metric associated with $x$ under consideration, which in this case is the detection metric (\ref{Eq:MLMS_Metric}). 
\item $f_x^{\mathtt{para}}(x)$ should be flexible enough to provide an accurate approximation of the exact distribution for a wide range of system parameters. 
\end{itemize}

For the problem at hand, the PDF of the CSI has to be supported only over the non-negative range, i.e., $f_{x}(x)=0$ for $x<0$, since  $\bar{c}_{\mathtt{s}}$ and $\bar{c}_{\mathtt{n}}$ assume only non-negative values. In addition, for the purposes of this paper, the adopted PDF $f^{\mathtt{para}}_{x}(x)$ should lead to a sufficiently simple detection metric (\ref{Eq:MLMS_Metric}). We have investigated several well-known  distributions as parametric models for the PDF of the CSI, including the Nakagami \cite{TableIntegSerie}, inverse Gaussian \cite{InvGaussian}, Levy  \cite{Levy}, and Gamma distributions \cite{TableIntegSerie}. We found that the Gamma distribution,  
\begin{align} \label{Eq:Gamma_PDF}
  f^{\mathtt{gamma}}_x(x) = \begin{cases}
  \frac{\beta^\alpha x^{\alpha-1} e^{-\beta x}}{\Gamma(\alpha)},\quad &\mathrm{if}\,\, x\geq 0 \\
  0, &\mathrm{otherwise}  
  \end{cases}
\end{align}
with parameters $\alpha,\beta>0$ \cite{TableIntegSerie}, is the best fit to the PDF of the CSI, $f_{x}(x),\,\,x\in\{\bar{c}_{\mathtt{s}},\bar{c}_{\mathtt{n}}\}$, for different system realizations. Moreover, it leads to a simple detection metric.  We emphasize that an \textit{experimentally} verified stochastic channel model for MC systems has not been reported yet. Despite this shortcoming, our motivation for adopting the Gamma distribution is as follows. First, for the stochastic MC channel used for simulation in Section~V,  the Gamma distribution can accurately model the randomness of the CSI introduced by random variations of the underlying MC channel parameters, e.g., the flow velocity, the enzyme concentration, the diffusion coefficient, etc. In fact, in Section~V, we show that the BERs of the optimal MS detector using the exact PDF of the CSI and the Gamma distribution perfectly match. Secondly, for the Gamma distribution, the ML decision metrics can be calculated in closed form. In particular, the expectations in the detection metric in  (\ref{Eq:MLMS_Metric}) are  of the form  $\mathbbmss{E}_{x}\left\{x^{a}e^{-bx}\right\}$ where $x\in\{\bar{c}_{\mathtt{s}},\bar{c}_{\mathtt{n}}\}$ and $a$ and $b$ are constants. Therefore, using the Gamma distribution,  $\mathbbmss{E}_{x}\left\{x^{a}e^{-bx}\right\}$ can be expressed as
\begin{align} \label{Eq:Metric_Gamma}
 \mathbbmss{E}_{x}\left\{x^{a}e^{-bx}\right\} & = \int_{x=0}^{\infty}  x^{a}e^{-bx} \times \frac{\beta^\alpha x^{\alpha-1} e^{-\beta x}}{\Gamma(\alpha)} \mathrm{d}x       
  =\frac{\beta^\alpha}{\Gamma(\alpha)} \int_{x=0}^{\infty} x^{a+\alpha-1}e^{-(b+\beta)x} \mathrm{d}x \nonumber \\
  & =\frac{\beta^\alpha \Gamma(a+\alpha)}{\Gamma(\alpha) (b+\beta)^{a+\alpha}} \underset{=1}{\underbrace{\int_{x=0}^{\infty} \frac{(b+\beta)^{a+\alpha} x^{a+\alpha-1}e^{-(b+\beta)x}}{\Gamma(a+\alpha)} \mathrm{d}x}}   = \frac{\Gamma(a+\alpha) \beta^\alpha }{\Gamma(\alpha) (b+\beta)^{a+\alpha}}.
\end{align}

The parameters $\alpha$ and $\beta$ have to be properly  chosen such that the resulting Gamma distribution well approximates the exact distribution or, if only the measurement data is available, the histogram of the measurement data. To this end, we adopt the weighted mean square error as a criterion to be minimized for the optimal choice of $\alpha$ and $\beta$. In particular, the optimal $\alpha^*$ and $\beta^*$ are obtained as
\begin{align} \label{Eq:Gamma_MMSE}
(\alpha^*,\beta^*) = 
\underset{\alpha,\beta>0}{\mathrm{argmin}}\,\,\int_{x=0}^{\infty} w(x) \left|f_x(x)-f_x^{\mathtt{gamma}}(x)\right|^2 \mathrm{d}x, \quad
\end{align}
where $w(x)\geq 0,\,\,\forall x$, is an appropriately chosen weight function that can be used to prioritize the accuracy of the approximation in a desired range of $x$. Using the Gamma distribution with the optimized parameters, the detection metrics required for the MS and SS detectors are given in closed form based on (\ref{Eq:Metric_Gamma}).

\begin{remk}\label{Remk:Search}
Since the feasible sets of $\alpha$ and $\beta$ are semi-infinite, i.e., $\alpha,\beta\in(0,+\infty)$, performing a full search to find the optimal $\alpha^*$ and $\beta^*$ is not possible. To overcome this challenge, we first note that for a Gamma distribution with mean $\mu_x$ and variance $\sigma_x^2$, the parameters $(\alpha,\beta)$ are \textit{uniquely} obtained~as 
\begin{align}\label{Eq:Gamma_Parameter}
(\alpha,\beta) = \left(\frac{\mu_x^2}{\sigma_x^2},\frac{\mu_x}{\sigma_x^2}\right).
\end{align}
Since the optimal parameters $(\alpha^*,\beta^*)$ from (\ref{Eq:Gamma_MMSE}) are expected to lead to a Gamma distribution that has a mean and a variance that are close to those of the exact distribution, we can efficiently bound the search for the optimal values to intervals $\alpha\in [(1-\delta)\frac{\bar{\mu}_x^2}{\bar{\sigma}_x^2},(1+\delta)\frac{\bar{\mu}_x^2}{\bar{\sigma}_x^2}]$ and $\beta\in [(1-\delta)\frac{\bar{\mu}_x}{\bar{\sigma}_x},(1+\delta)\frac{\bar{\mu}_x}{\bar{\sigma}_x}]$, where $\bar{\mu}_x$ and $\bar{\sigma}_x^2$ are the mean and the variance of the exact distribution, respectively, and $\delta\geq 0$ determines how large the search intervals are.  
\end{remk}

\subsection{Non-Coherent DF Detector}

As will be shown in Section~V, the proposed optimal MS detector significantly outperforms the optimal SS detector, particularly when the number of jointly detected symbols is large. However, the gain obtained by the optimal MS detector in this case comes at the expense of a large detection delay. In particular, we have to wait until all $K$ symbols in one detection window arrive before detection can be performed. In order to mitigate this problem and to further reduce complexity while still exploiting the benefits of MS detection, we consider a DF detector in the following. In particular, the DF detector  instantly detects the received symbols while exploiting the memory of the previously detected symbols assuming that these symbols were \textit{correctly} detected. 

To characterize the DF detector, let us first define $\mathbf{r}_k=[r[k-1],\dots,r[k-K+1]]^T$ and $\hat{\mathbf{s}}_k=[\hat{s}[k-1],\hat{s}[k-2],\dots,\hat{s}[k-K+1]]^T$ as the observation vector and the detected symbol vector for the previous $K-1$ symbol intervals.  The  DF detector is derived in a similar manner as the optimal MS detector, cf. Appendix~\ref{App:Theo_MLMS}. In particular, given the $K-1$ previously detected symbols in $\hat{\mathbf{s}}_k$, we  evaluate the detection metric in (\ref{Eq:MLMS_Metric}) for the two possibilities $s[k]=1$ and $s[k]=0$ which leads to  the detection metrics $\Lambda^{\mathtt{MS}}(\mathbf{s}_1,\mathbf{r}) $ and $ \Lambda^{\mathtt{MS}}(\mathbf{s}_0,\mathbf{r})$, respectively, where  $\mathbf{s}_0=[0,\hat{\mathbf{s}}_k^T]^T$, $\mathbf{s}_1=[1,\hat{\mathbf{s}}_k^T]^T$, and $\mathbf{r}=[r[k],\mathbf{r}_k^T]^T$. This leads to the following decision rule.

\textit{Proposed DF Detector:}
The non-coherent DF detector for the MC system considered in this paper is given~by 
\begin{IEEEeqnarray}{lll} \label{Eq:MLDF_sol_metric}
  \hat{s}^{\mathtt{DF}}[k] = \begin{cases}
  1,\quad &\mathrm{if}\,\,  \Lambda^{\mathtt{MS}}(\mathbf{s}_1,\mathbf{r}) \geq \Lambda^{\mathtt{MS}}(\mathbf{s}_0,\mathbf{r}) \\
  0,\quad &\mathrm{otherwise} \\
  \end{cases}
\end{IEEEeqnarray}
Since the above DF detector uses the MS detection metric in (\ref{Eq:MLMS_Metric}), we can also employ (\ref{Eq:Metric_Gamma}) to obtain an approximate metric based on the Gamma distribution proposed in Subsection~III-A. 
 
Given $\mathbf{r}_k$ and $\hat{\mathbf{s}}_k$, the ratio $\Lambda^{\mathtt{MS}}(\mathbf{s}_1,\mathbf{r}) / \Lambda^{\mathtt{MS}}(\mathbf{s}_0,\mathbf{r})$ is a monotonically increasing function of $r[k]$. Since $\Lambda^{\mathtt{MS}}(\mathbf{s}_1,\mathbf{r}) < \Lambda^{\mathtt{MS}}(\mathbf{s}_0,\mathbf{r})$ holds for  $r[k]=0$, there exists a unique threshold for $r[k]$, denoted by $\xi^{\mathtt{DF}}(\hat{\mathbf{s}}_k,\mathbf{r}_k)$, above which $\Lambda^{\mathtt{MS}}(\mathbf{s}_1,\mathbf{r}) > \Lambda^{\mathtt{MS}}(\mathbf{s}_0,\mathbf{r})$ holds. Hence, the non-coherent DF detector can  be equivalently written in the following threshold-based form
\begin{IEEEeqnarray}{lll} \label{Eq:MLDF_sol_thr}
  \hat{s}^{\mathtt{DF}}[k] = \begin{cases}
  1,\quad &\mathrm{if}\,\,  r[k]\geq \xi^{\mathtt{DF}}(\hat{\mathbf{s}}_k,\mathbf{r}_k) \\
  0,\quad &\mathrm{otherwise} \\
  \end{cases}
\end{IEEEeqnarray}
The adaptive DF detection threshold $\xi^{\mathtt{DF}}(\hat{\mathbf{s}}_k,\mathbf{r}_k)$ is given by
\begin{IEEEeqnarray}{lll} \label{Eq:MLDF_thr}
\xi^{\mathtt{DF}}(\hat{\mathbf{s}}_k,\mathbf{r}_k) 
 =\min\Big\{\xi\in\mathbb{N}\big| \Lambda^{\mathtt{MS}}(\mathbf{s}_1,\mathbf{r}^\xi) > \Lambda^{\mathtt{MS}}(\mathbf{s}_0,\mathbf{r}^\xi)  \Big\},
\end{IEEEeqnarray}
where $\mathbf{r}^\xi=[\xi,\mathbf{r}_k^T]^T$. We will employ the threshold detection form of the proposed DF detector in (\ref{Eq:MLDF_sol_thr}) for the performance analysis provided in Section~IV-C.


\subsection{Suboptimal Detector Based on Blind CSI Estimation}

The non-coherent detectors developed so far require \textit{statistical}  knowledge of the CSI which might be difficult to acquire for some  MC systems deployed in practice. Therefore, in the following, we  propose a suboptimal detector that does not need  statistical CSI knowledge. The main idea behind the simple detector, which we propose in this subsection, is to first estimate the CSI based on the   symbols received in the considered detection window in order to approximate the optimal ML threshold which is denoted by $\xi^{\mathtt{ML}}_{\mathtt{BL}}$. Subsequently, symbol-by-symbol detection can be performed based on the approximated threshold $\xi^{\mathtt{ML}}_{\mathtt{BL}}$. We note that the channel estimator is blind in the sense that a training sequence is not used.

 For a given observation block $\mathbf{r}$, let  $\widetilde{\mathcal{K}}_1$ ($\widetilde{\mathcal{K}}_0$) denote the sets of indices $k$ for the $\left\lceil \frac{K}{2} \right\rceil$-th largest ($\left\lfloor \frac{K}{2} \right\rfloor$-th smallest) $r[k]$  in the block. The proposed suboptimal detector is formally presented in the following. 

\textit{Proposed Blind ML-Based Detector:} The  proposed blind detector for  ON-OFF keying modulation in diffusive MC is given by
\begin{align} \label{Eq:MLBL_sol}
  \hat{s}^{\mathtt{ML}}_{\mathtt{BL}}[k] = \begin{cases}
  1,\quad &\mathrm{if}\,\, r[k] \geq \xi^{\mathtt{ML}}_{\mathtt{BL}}(\mathbf{r}) \\
  0,\quad &\mathrm{otherwise} \\
  \end{cases}
\end{align}
where  $\xi^{\mathtt{ML}}_{\mathtt{BL}}(\mathbf{r}) = \frac{\hat{\bar{c}}_{\mathtt{s}}}{\mathrm{ln}(1+\frac{\hat{\bar{c}}_{\mathtt{s}}}{\hat{\bar{c}}_{\mathtt{n}}})}$ is the detection threshold. Hereby, the
CSI estimates $\hat{\bar{c}}_{\mathtt{s}}$ and $\hat{\bar{c}}_{\mathtt{n}}$ are obtained as
\begin{subequations}\label{Eq:CIR_Est}
\begin{align} 
   \hat{\bar{c}}_{\mathtt{s}} &= \frac{1}{\left\lceil \frac{K}{2} \right\rceil} \sum_{k\in\widetilde{\mathcal{K}}_1} [r[k] - \hat{\bar{c}}_{\mathtt{n}} ]  \label{Eq:CIR_Est_a} \\
    \hat{\bar{c}}_{\mathtt{n}} &= \frac{1}{\left\lfloor \frac{K}{2} \right\rfloor} \sum_{k\in\widetilde{\mathcal{K}}_0} r[k].   \label{Eq:CIR_Est_b}
\end{align}
\end{subequations}

The CSI estimates $\hat{\bar{c}}_{\mathtt{s}}$ and $\hat{\bar{c}}_{\mathtt{n}}$ in (\ref{Eq:CIR_Est}) correspond to a simple averaging over the expected positions of $s[k]=1$ and $s[k]=0$, respectively. Since the noise molecules are always present whereas the signal molecules are present only when $s[k]=1$ holds, the \textit{expected} number of molecules observed at the receiver in positions with $s[k]=1$ is  larger than that in positions with $s[k]=0$.  Therefore, we first compute $\hat{\bar{c}}_{\mathtt{n}}$ directly from (\ref{Eq:CIR_Est_b}) and then  $\hat{\bar{c}}_{\mathtt{s}}$ from (\ref{Eq:CIR_Est_a}). Having the estimated CSI $(\hat{\bar{c}}_{\mathtt{s}},\hat{\bar{c}}_{\mathtt{n}})$, we can compute the ML threshold $\xi^{\mathtt{ML}}_{\mathtt{BL}}$ and perform detection based on (\ref{Eq:MLSS_sol}). We note that although $s[k]=1$ and $s[k]=0$ are equiprobable, the number of ones and zeros in each detection window may not be exactly $\frac{K}{2}$. Therefore, assuming that the $\left\lceil \frac{K}{2}\right\rceil$-th  largest elements ($\left\lfloor \frac{K}{2} \right\rfloor$-th smallest elements) of $\mathbf{r}$ correspond to the positions of $s[k]=1$ ($s[k]=0$) leads to an inherent CSI estimation error.

\section{Performance Analysis}

In this section, we derive exact expressions and  performance bounds for the BERs of the proposed detection schemes. In particular, the average BER, denoted by $\bar{P}_e$, can be written mathematically as
\begin{IEEEeqnarray}{lll} \label{Eq:BER_Avg}
\bar{P}_e = \mathbbmss{E}_{\bar{\mathbf{c}}}\big\{P_e(\bar{\mathbf{c}})\big\} = \int_{\bar{c}_{\mathtt{s}}\geq 0} \int_{\bar{c}_{\mathtt{n}}\geq 0} \Pr\big\{\hat{s}[k]\neq s[k]|\bar{\mathbf{c}}\big\}  f_{\bar{c}_{\mathtt{s}}}(\bar{c}_{\mathtt{s}})  f_{\bar{c}_{\mathtt{n}}}(\bar{c}_{\mathtt{n}}) \mathrm{d}\bar{c}_{\mathtt{s}}\mathrm{d}\bar{c}_{\mathtt{n}},
\end{IEEEeqnarray}
where $P_e(\bar{\mathbf{c}})\triangleq\Pr\big\{\hat{s}[k]\neq s[k]|\bar{\mathbf{c}}\big\}$ is the  BER conditioned on a particular state of the MC channel characterized by the CSI $\bar{\mathbf{c}}$. In the following, we derive the conditional BER of the SS detector and conditional bounds for the MS detector.

\subsection{BER of the Optimal SS Detector}

We first derive the exact expression for the BER of the optimal SS detector. In particular, for a given $\bar{\mathbf{c}}$, the BER of the optimal SS detector is obtained as
\begin{IEEEeqnarray}{rll} \label{Eq:BER_MLSS}
P_e^{\mathtt{SS}}(\bar{\mathbf{c}}) = \,\,& 
\Pr\big\{\hat{s}[k]=0|\bar{\mathbf{c}},s[k]=1\big\}\Pr\big\{s[k]=1\big\}  \nonumber \\
& + \Pr\big\{\hat{s}[k]=1|\bar{\mathbf{c}},s[k]=0\big\}\Pr\big\{s[k]=0\big\} \nonumber \\
= \,\,&  \frac{1}{2} \Pr\big\{r[k] < \xi^{\mathtt{SS}}|\bar{\mathbf{c}},s[k]=1\big\}
+  \frac{1}{2} \Pr\big\{r[k]\geq \xi^{\mathtt{SS}}|\bar{\mathbf{c}},s[k]=0\big\}\quad \nonumber \\
\overset{(a)}{=}\,\,& \frac{1}{2} \sum_{r[k]=0}^{\lfloor\xi^{\mathtt{SS}}\rfloor}  \frac{\left( \bar{c}_{\mathtt{s}} + \bar{c}_{\mathtt{n}} \right)^{r[k]} \exp\left(- \bar{c}_{\mathtt{s}} - \bar{c}_{\mathtt{n}} \right)}{r[k]!}  
+ \frac{1}{2} \sum_{r[k]=\lceil\xi^{\mathtt{SS}}\rceil}^{\infty} \frac{\bar{c}_{\mathtt{n}}^{r[k]} \exp\left(- \bar{c}_{\mathtt{n}} \right)}{r[k]!} \nonumber \\
\overset{(b)}{=}\,\,& \frac{1}{2} + \frac{1}{2} \Big[ Q\left(\lceil \xi^{\mathtt{SS}} \rceil, \bar{c}_{\mathtt{s}} + \bar{c}_{\mathtt{n}} \right)
-Q\left(\lceil \xi^{\mathtt{SS}} \rceil, \bar{c}_{\mathtt{n}} \right) \Big],
\end{IEEEeqnarray}
where $Q(x,y)=\frac{\Gamma(x,y)}{\Gamma(x)}$ is the regularized Gamma function and $\Gamma(\cdot,\cdot)$ is the upper incomplete Gamma function \cite{TableIntegSerie}. For equality $(a)$, we used the fact that $r[k]$ conditioned on $\bar{\mathbf{c}}$ and $s[k]$ follows a Poisson distribution, and for $(b)$, we exploited the expression for the CDF of a Poisson RV with mean $\lambda$ given by $F_{r[k]}(\xi)=\Pr\{r[k]\leq\xi\}=Q(\lfloor \xi+1\rfloor,\lambda)$ which leads to $\Pr\{r[k]<\xi\}=Q(\lceil \xi\rceil,\lambda)$ \cite{TableIntegSerie}.

\begin{remk}
The BER expression in (\ref{Eq:BER_MLSS}) can be also used to compute the BER of the benchmark scheme with perfect CSI knowledge given in Subsection~III.A after replacing $\xi^{\mathtt{SS}}$ with $\xi^{\mathtt{ML}}(\bar{\mathbf{c}})$.   
\end{remk}

\subsection{BER Upper Bound for the Optimal MS  Detector}

Unfortunately, the derivation of an exact expression for the BER of the optimal MS detector is a difficult, if
not impossible,  task due to the multi-dimensional structure of the statistics of the observation vector. Hence, in this subsection, we use the union bound to arrive at an upper bound on the BER of the optimal MS detector.

The pairwise error probability (PEP), denoted by $P(\mathbf{s}\to\hat{\mathbf{s}})$, is defined as the probability that, assuming $\mathbf{s}$ is transmitted, $\hat{\mathbf{s}}$ is detected. We note that, due to the structure of the optimal MS detector, the error probabilities for all  transmitted sequences which have the same number of ones are identical. Hence, without loss of generality, we consider only $K+1$ sequences $\mathbf{s}$ whose first $n_1\in\{0,\dots,K\}$, bits are ones and which are collected in set $\tilde{\mathcal{A}}$. Using the PEP, the BER is upper bounded based on the union bound as follows
\begin{IEEEeqnarray}{rll} \label{Eq:PEP_r}
P_e^{\mathtt{MS}}(\bar{\mathbf{c}}) \,\,& \leq \frac{1}{K} \sum_{\mathbf{s}\in\hat{\mathcal{A}}} \sum_{\hat{\mathbf{s}}\in\{\mathcal{A}\backslash\mathbf{s}\}} h\left(\mathbf{s},\hat{\mathbf{s}}\right) P^{\mathtt{MS}}(\mathbf{s}\to\hat{\mathbf{s}}) \Pr\{\mathbf{s}\},
\end{IEEEeqnarray}
where $h\left(\mathbf{s},\hat{\mathbf{s}}\right)$ denotes the Hamming distance between $\mathbf{s}$ and $\hat{\mathbf{s}}$ and $\Pr\{\mathbf{s}\}=\frac{1}{2^K}{K\choose n_1}$ where $n_1$ is the number of ones in $\mathbf{s}$.
We note that in order to calculate the PEP for a given $\mathbf{s}$ and $\hat{\mathbf{s}}$, a $K$-dimensional summation with respect to the different possibilities for the observation vector $\mathbf{r}$ is needed, which is a computationally challenging task.  However, as can be seen from (\ref{Eq:MLMS_Metric}), the detection metric is a function  solely of $n_1$, $N_1$, and $N=N_1+N_0$. Thus, we introduce the new notation $\Lambda^{\mathtt{MS}}(n_1,N_1,N)$ to indicate this. Let $\Lambda^{\mathtt{MS}}(n_1,N_1,N)$ and $\Lambda^{\mathtt{MS}}(\hat{n}_1,\hat{N}_1,N)$ be the detection metrics for hypothesis sequences $\mathbf{s}$ and $\hat{\mathbf{s}}$, respectively, for a  given $\mathbf{r}$.
We note that, given $\mathbf{s}$ and $\hat{\mathbf{s}}$, the values of $n_1$ and $\hat{n}_1$ are known, respectively, and for a given $\mathbf{r}$, the value of $N$ is the same for both $\mathbf{s}$ and $\hat{\mathbf{s}}$. Hence, unlike the expression in (\ref{Eq:PEP_r}), which requires a $K$-dimensional summation with respect to the elements of $\mathbf{r}$,  it suffices to consider a three-dimensional summation with respect to $N_1$, $\hat{N}_1$, and $N$. In particular,  in order to calculate $P^{\mathtt{MS}}(\mathbf{s}\to\hat{\mathbf{s}})$, we have to find the probabilities of  the observation events $(N_1,\hat{N}_1,N)$, denoted by $\Pr\big\{ N_1,\hat{N}_1,N | \bar{\mathbf{c}}\big\}$, for which $\Lambda^{\mathtt{MS}}(n_1,N_1,N)<\Lambda^{\mathtt{MS}}(\hat{n}_1,\hat{N}_1,N)$ holds. This leads to
\begin{IEEEeqnarray}{rll}  \label{Eq:PEP_N}
P^{\mathtt{MS}}(\mathbf{s}\to\hat{\mathbf{s}}) &= \Pr\left\{\Lambda^{\mathtt{MS}}(\mathbf{s},\mathbf{r})<\Lambda^{\mathtt{MS}}(\hat{\mathbf{s}},\mathbf{r})\right\} \nonumber \\
&=\sum_{N}\sum_{N_1}\sum_{\hat{N}_1}\mathbf{1}\big\{\Lambda^{\mathtt{MS}}(n_1,N_1,N)<\Lambda^{\mathtt{MS}}(\hat{n}_1,\hat{N}_1,N)\big\} \Pr\big\{ N_1,\hat{N}_1,N | \bar{\mathbf{c}}\big\}.\quad
\end{IEEEeqnarray}

Conditioned on $\bar{\mathbf{c}}$, variables $N_1$, $\hat{N}_1$, and $N$ are \textit{correlated} Poisson RVs. Hence, in order to compute $\Pr\big\{ N_1,\hat{N}_1,N| \bar{\mathbf{c}} \big\}$, we divide the observation vector $\mathbf{r}$ into the following four partitions:

$\mathcal{S}_1$: Positions of $\mathbf{r}$ where the corresponding elements of both $\mathbf{s}$ and $\hat{\mathbf{s}}$ are one.

$\mathcal{S}_2$: Positions of $\mathbf{r}$ where the corresponding elements of $\mathbf{s}$ and $\hat{\mathbf{s}}$ are one and zero, respectively.

$\mathcal{S}_3$: Positions of $\mathbf{r}$ where the corresponding elements of $\mathbf{s}$ and $\hat{\mathbf{s}}$ are zero and one, respectively.

$\mathcal{S}_4$: Positions of $\mathbf{r}$ where the corresponding elements of both $\mathbf{s}$ and $\hat{\mathbf{s}}$ are zero.

\noindent 
The sum of observed molecules in the positions specified by $\mathcal{S}_1$, $\mathcal{S}_2$, $\mathcal{S}_3$, and $\mathcal{S}_4$ are denoted by $M_1$, $M_2$, $M_3$, and $M_4$, respectively. It follows that, conditioned on $\bar{\mathbf{c}}$, variables $M_1$, $M_2$, $M_3$, and $M_4$ are \textit{independent} Poisson RVs with means $\lambda_1=\upsilon(\bar{c}_{\mathtt{s}} + \bar{c}_{\mathtt{n}})$, $\lambda_2=(n_1-\upsilon)(\bar{c}_{\mathtt{s}} + \bar{c}_{\mathtt{n}})$, $\lambda_3=(\hat{n}_1-\upsilon)  \bar{c}_{\mathtt{n}}$, and $\lambda_4=(K+\upsilon-n_1-\hat{n}_1)\bar{c}_{\mathtt{n}}$, respectively, where $\upsilon=|\mathcal{S}_1|$. Moreover, we have $N_1=M_1+M_2$, $\hat{N}_1=M_1+M_3$, and $N=M_1+M_2+M_3+M_4$.  Using these results, we obtain
\begin{IEEEeqnarray}{lll} \label{Eq:PEP_ind} 
\Pr \big\{ N_1,\hat{N}_1,N | \bar{\mathbf{c}}\big\} \,\,& =  \sum_{i=0}^{\min\{N_1,\hat{N}_1\}} \Pr\big\{ M_1=i | \bar{\mathbf{c}}\big\} \Pr \big\{M_2=N_1-i | \bar{\mathbf{c}}\big\} \nonumber \\
&\qquad\qquad\quad \times\Pr \big\{M_3=\hat{N}_1-i| \bar{\mathbf{c}}\big\} \Pr \big\{M_4=N_1+\hat{N}_1-i| \bar{\mathbf{c}}\big\}\nonumber \\
&\,\, = \sum_{i=0}^{\min\{N_1,\hat{N}_1\}} \frac{\lambda_1^{i}e^{-i\lambda_1}}{i!} \times\frac{\lambda_2^{N_1-i}e^{-(N_1-i)\lambda_2}}{(N_1-i)!} 
\nonumber \\
&\qquad\qquad\quad  \times\frac{\lambda_3^{\hat{N}_1-i}e^{-(\hat{N}_1-i)\lambda_3}}{(\hat{N}_1-i)!} \times\frac{\lambda_4^{N_1+\hat{N}_1-i}e^{-(N_1+\hat{N}_1-i)\lambda_4}}{(N_1+\hat{N}_1-i)!}.
\end{IEEEeqnarray}
Using (\ref{Eq:BER_Avg}) and (\ref{Eq:PEP_r})-(\ref{Eq:PEP_ind}), the average PEP can be analytically computed. 

\begin{remk}
In Section~V, we employ (\ref{Eq:PEP_r}) as an upper bound for the BER of the optimal MS detector and verify its tightness by comparing it with the exact BER obtained via simulation. We note that the computation of the \textit{average} PEP in an analytical form based on (\ref{Eq:BER_Avg}) and (\ref{Eq:PEP_r})-(\ref{Eq:PEP_ind}) is a computationally challenging task due to the numerical integration required for evaluating (\ref{Eq:BER_Avg}). In order to reduce the complexity of the evaluation of the average PEP, one can employ a hybrid approach where, for a given CSI $\bar{\mathbf{c}}$, the PEP bound in (\ref{Eq:PEP_r}) is calculated numerically and the expectation in (\ref{Eq:BER_Avg}) is performed via Monte Carlo simulation. 
\end{remk}

\subsection{BER Lower Bound for the MS/DF Detectors}

In this subsection, we derive the BER of a genie-aided DF detector, i.e., we assume \textit{error-free} decision feedback, which  constitutes  a lower bound on the BERs of the MS/DF detectors. Similar to the MS detection metric,  the DF detection threshold $\xi^{\mathtt{DF}}(\hat{\mathbf{s}}_k,\mathbf{r}_k)$ is also a function of only $n_{k,1}$, $N_{k,1}$, $N_{k,0}$, i.e., we may use the notation $\xi^{\mathtt{DF}}(n_{k,1},N_{k,1},N_{k,0})$. Exploiting this observation, the BER of the genie-aided DF detector conditioned on $\bar{\mathbf{c}}$ is written as 
\begin{IEEEeqnarray}{rll} \label{Eq:BER_MLDF}
P_e^{\mathtt{DF}}(\bar{\mathbf{c}}) & =  
\sum_{N_{k,0}}\sum_{N_{k,1}}\sum_{n_{k,1}}\Pr\big\{\hat{s}[k]\neq s[k]|\bar{\mathbf{c}},n_{k,1},N_{k,1},N_{k,0}\big\}\Pr\big\{n_{k,1},N_{k,1},N_{k,0}|\bar{\mathbf{c}}\big\} \nonumber \\
& = \sum_{N_{k,0}}\sum_{N_{k,1}}\sum_{n_{k,1}} \Pr\big\{\hat{s}[k]\neq s[k]|\bar{\mathbf{c}},n_{k,1},N_{k,1},N_{k,0}\big\} \Pr\big\{N_{k,1},N_{k,0}|\bar{\mathbf{c}},n_{k,1}\big\} 
\Pr\big\{n_{k,1}|\bar{\mathbf{c}}\big\}.\quad\,\,
\end{IEEEeqnarray}
In (\ref{Eq:BER_MLDF}), $n_{k,1}$  denotes the number of ones in the $K-1$ previously detected symbols and is a binomial RV. Moreover, $N_{k,1}$ and $N_{k,0}$ conditioned on $\bar{\mathbf{c}}$ and $n_{k,1}$ are independent Poisson RVs with means $n_{k,1}(\bar{c}_{\mathtt{s}}+\bar{c}_{\mathtt{n}})$ and $(K-n_{k,1})\bar{c}_{\mathtt{n}}$, respectively. Therefore, $\Pr\big\{n_{k,1}|\bar{\mathbf{c}}\big\}$,  $\Pr\big\{N_{k,1}|\bar{\mathbf{c}},n_{k,1}\big\}$, and $\Pr\big\{N_{k,0}|\bar{\mathbf{c}},n_{k,1}\big\}$ can be computed as
\begin{IEEEeqnarray}{rll} \label{Eq:BER_MLDF_Prob}
\Pr\big\{n_{k,1}|\bar{\mathbf{c}}\big\} &\,\,=  \frac{1}{2^{K-1}} {K-1 \choose n_{k,1}} \IEEEyesnumber\IEEEyessubnumber\\
\Pr\big\{N_{k,1}|\bar{\mathbf{c}},n_{k,1}\big\} &\,\,= \frac{n_{k,1}^{N_{k,1}}(\bar{c}_{\mathtt{s}}+\bar{c}_{\mathtt{n}})^{N_{k,1}} e^{-n_{k,1}(\bar{c}_{\mathtt{s}}+\bar{c}_{\mathtt{n}})}}{N_{k,1}!}\IEEEyessubnumber \\
\Pr\big\{N_{k,0}|\bar{\mathbf{c}},n_{k,1}\big\} &\,\,= \frac{(K-n_{k,1})^{N_{k,0}}\bar{c}_{\mathtt{n}}^{N_{k,0}} e^{-(K-n_{k,1})\bar{c}_{\mathtt{n}}}}{N_{k,0}!}.\IEEEyessubnumber
\end{IEEEeqnarray}
Furthermore, $\Pr\big\{\hat{s}[k]\neq s[k]|\bar{\mathbf{c}},n_{k,1},N_{k,1},N_{k,0}\big\}$ for the genie-aided DF detector is given by
\begin{IEEEeqnarray}{lll} \label{Eq:BER_MLDF3}
\Pr\big\{\hat{s}[k]\neq s[k]|\bar{\mathbf{c}},n_{k,1},N_{k,1},N_{k,0}\big\} 
 & =    \frac{1}{2} \Pr\big\{r[k] < \xi^{\mathtt{DF}}(n_{k,1},N_{k,1},N_{k,0})|\bar{\mathbf{c}},s[k]=1\big\} \nonumber \\
  &\quad +  \frac{1}{2} \Pr\big\{r[k]\geq \xi^{\mathtt{DF}}(n_{k,1},N_{k,1},N_{k,0})|\bar{\mathbf{c}},s[k]=0\big\}\quad \nonumber \\
& = \frac{1}{2} + \frac{1}{2} \Big[ Q\left(\lceil \xi^{\mathtt{DF}}(n_{k,1},N_{k,1},N_{k,0}) \rceil, \bar{c}_{\mathtt{s}} + \bar{c}_{\mathtt{n}} \right)\nonumber \\
&\quad -Q\left(\lceil \xi^{\mathtt{DF}}(n_{k,1},N_{k,1},N_{k,0}) \rceil, \bar{c}_{\mathtt{n}} \right) \Big].
\end{IEEEeqnarray}

\section{Numerical Results}

In this section, we first present the stochastic MC channel model and the adopted system parameters used for simulation. Subsequently, we evaluate the performance of the  proposed non-coherent detectors where we employ the optimal coherent detector as a benchmark scheme.

\subsection{Stochastic MC Channel Model}

In this section, we present the stochastic MC channel model used for the simulation results. However, we emphasize that the proposed non-coherent detection framework is valid for any given $f_{\bar{c}_{\mathtt{s}}}(\bar{c}_{\mathtt{s}}) $ and $ f_{\bar{c}_{\mathtt{n}}}(\bar{c}_{\mathtt{n}})$ and is not limited to the stochastic MC channel model used here for simulation.

Let us assume a point source with impulsive molecule release, a fully transparent spherical receiver with volume $V^{\mathtt{RX}}$,  and an unbounded environment with diffusion coefficient $D$. Moreover, we denote the distance between the transmitter and the receiver by $d$. In addition, we assume that  there is steady uniform flow (or drift)  with parallel and perpendicular  velocity components, denoted by $v_{\parallel}$ and $v_{\perp}$, respectively, with respect to the direction from the transmitter to the receiver. Furthermore, the signaling molecules may react with  enzyme molecules, which are present in the MC environment, and degrade into a form that cannot be detected by the receiver. We assume a uniform and constant concentration of the enzyme, denoted by $\bar{c}_{\mathtt{e}}$, and a first order reaction mechanism between the signaling and enzyme molecules with constant reaction rate $\kappa$ \cite{Adam_OptReciever}.  Based on the aforementioned assumptions,  the \textit{expected} number of molecules observed at the receiver as a function of time, denoted by $\bar{c}_{\mathtt{s}}(t)$, is given by \cite{Adam_OptReciever}
\begin{align} \label{Eq:Consentration}
  \bar{c}_{\mathtt{s}}(t) = \frac{N^{\mathtt{TX}}V^{\mathtt{RX}}}{(4\pi D t)^{3/2}} \mathrm{exp}\left(-\kappa\bar{c}_{\mathtt{e}}t-\frac{(d-v_{\parallel}t)^2+(v_{\perp}t)^2}{4Dt}\right). \quad  \hspace{-0.5cm}
\end{align}
Furthermore, assuming a peak observation detector, the sample time from the beginning of each symbol interval is chosen as $t^{\mathtt{max}}=\underset{t>0}{\mathrm{argmax}}\,\,\bar{c}_{\mathtt{s}}(t)$ which leads to $\bar{c}_{\mathtt{s}}=\underset{t>0}{\mathrm{max}}\,\,\bar{c}_{\mathtt{s}}(t)$.

The channel parameters undergo random variations that lead to random variations in $\bar{c}_{\mathtt{s}}$. For instance, the flow velocity components, $v_{\parallel}$ and $v_{\perp}$, may vary over time or the diffusion coefficient, $D$, and enzyme concentration, $\bar{c}_{\mathtt{e}}$, may change due to variations in the environment temperature. To capture these effects,  we assume that the channel parameters  in each detection window are realizations of RVs according to  $z=z^{\mathtt{def}}(1+\sigma_{z}\mathcal{N}\left(0,1\right)),\,\,z\in\{D,v_{\parallel},v_{\perp},\bar{c}_{\mathtt{e}}\}$ where $z^{\mathtt{def}}$ denotes the mean value of parameter $z$ and $\sigma_{z}z^{\mathtt{def}}$ is its standard deviation, which determines how much the parameter may deviate from the mean\footnote{We note that a Gaussian RV may assume negative values whereas $D$ and $\bar{c}_{\mathtt{e}}$  are non-negative parameters. Therefore, we assume small values for $\sigma_{D}$ and $\sigma_{\bar{c}_{\mathtt{e}}}$,  and omit those realizations for which $z\in\{D,\bar{c}_{\mathtt{e}}\}$ is negative.}. As $\sigma_z\to 0,\,\,\forall z$, the respective MC channel becomes deterministic, and for large  $\sigma_z$, the corresponding MC channel is highly stochastic. Substituting the Gaussian RVs $z\in\{D,v_{\parallel},v_{\perp},\bar{c}_{\mathtt{e}}\}$ into (\ref{Eq:Consentration}) may not lead to a closed-form analytical expression for $f_{\bar{c}_{\mathtt{s}}}(\bar{c}_{\mathtt{s}}) $. Therefore, we employ Monte Carlo simulation to determine a histogram for $\bar{c}_{\mathtt{s}}$ as the ``true" distribution, $f_{\bar{c}_{\mathtt{s}}}(\bar{c}_{\mathtt{s}})$, and to verify the effectiveness of the proposed Gamma distribution to approximate the true distribution. For the results provided in this section,  we assume that the variation of the  mean of the noise $\bar{c}_{\mathtt{n}}$ is modeled similarly  to the variation of the mean of the signal $\bar{c}_{\mathtt{s}}$. In particular, we choose $\bar{c}_{\mathtt{n}}=\mathtt{SNR}^{-1} X$ where $X$ is an RV whose PDF is identical to that of $\bar{c}_{\mathtt{s}}$ and $\mathtt{SNR}$ is a constant analogous to the signal-to-noise ratio (SNR) in conventional wireless systems. Furthermore, if $\bar{c}_{\mathtt{s}}\sim\mathrm{Gamma}(\alpha,\beta)$ holds, we obtain $\bar{c}_{\mathtt{n}}\sim\mathrm{Gamma}(\alpha,\mathtt{SNR}\beta)$ \cite{TableIntegSerie}. 

\begin{table}
\label{Table:Parameter}
\caption{Default Values of the Numerical Parameters \cite{Adam_OptReciever,CellBio}. \vspace{-0.3cm}} 
\begin{center}
\scalebox{0.59}
{
\begin{tabular}{|| c | c | c ||}
  \hline 
  Variable & Definition & Value \\ \hline \hline
       $V^{\mathtt{RX}}$ & Receiver volume   & $\frac{4}{3}\pi 50^3$ \,\, ${\text{nm}}^3$ \\  
         &    & (a sphere with radius $50$ nm) \\ \hline   
        $d$ &  Distance between the transmitter and the receiver  & $500$ nm\\ \hline 
         $D$ &  Diffusion coefficient for the signaling molecule & $4.365\times 10^{-10}$ $\text{m}^2\cdot\text{s}^{-1}$\\ \hline          
       $\bar{c}_{\mathtt{e}}$ &  Enzyme concentration  & $10^{5}$ $\text{molecule}\cdot\mu\text{m}^3$ \\   
         &     & (approx. $1.66$ micromolar) \\ \hline  
       $\kappa$ &    Rate of molecule degradation reaction & $2\times10^{-19}$ $\text{m}^3\cdot\text{molecule}^{-1}\cdot\text{s}^{-1}$ \\ \hline 
       $(v_{\parallel},v_{\perp})$ &  Components of flow velocity   & $(10^{-3},10^{-3})$ $\text{m}\cdot\text{s}^{-1}$ \\ \hline
\end{tabular}
}
\end{center}\vspace{-0.3cm}
\end{table}

\subsection{Simulation Parameters, DF Window Size, and Benchmark Scheme}

\textit{Simulation Parameters:} The default values of the channel parameters are given in Table~I.  Moreover, we consider the following three scenarios for the stochastic MC channel: Scenario~1: $(\sigma_{D},\sigma_{v_{\parallel}},\sigma_{v_{\perp}},\sigma_{\bar{c}_{\mathtt{e}}})=(0.1,0.5,0.5,0.1)$, Scenario~2: $(\sigma_{D},\sigma_{v_{\parallel}},\sigma_{v_{\perp}},\sigma_{\bar{c}_{\mathtt{e}}})=(0.2,1,1.5,0.1)$, and Scenario~3: $(\sigma_{D},\sigma_{v_{\parallel}},\sigma_{v_{\perp}},\sigma_{\bar{c}_{\mathtt{e}}})=(0.1,1.5,0.5,0.2)$.  Each of these scenarios leads to a pair of PDFs for  $\bar{c}_{\mathtt{s}}$ and $\bar{c}_{\mathtt{n}}$. In particular, we choose $\bar{c}_{\mathtt{n}}$ such that for the default parameters in Table~I and $N^{\mathtt{Tx}}=10^{4}$, we obtain $\mathbbmss{E}\{\bar{c}_{\mathtt{n}}\}=\mathbbmss{E}\{\bar{c}_{\mathtt{s}}\}$, i.e., $\mathtt{SNR}=1$ holds for this case. To obtain different SNRs, we change the number of molecules released by the transmitter. We used Monte Carlo simulation to obtain simulation results. Thereby, we first generated $N=10^6$ realizations of the CSI vector $\bar{\mathbf{c}}$. Then, for each $\bar{\mathbf{c}}$, we generated the number of  molecules observed at the receiver in each symbol interval based on the channel model in (\ref{Eq:ChannelInOut}) where RVs $c_{\mathtt{s}}[k] $ and $ c_{\mathtt{n}}[k]$ are Poisson distributed with means $\bar{c}_{\mathtt{s}}$ and $\bar{c}_{\mathtt{n}}$, respectively. For the analytical results, we numerically evaluate the expression~in~(\ref{Eq:BER_Avg}).

\begin{figure}
  \centering
 \scalebox{0.8}{
\pstool[width=0.8\linewidth]{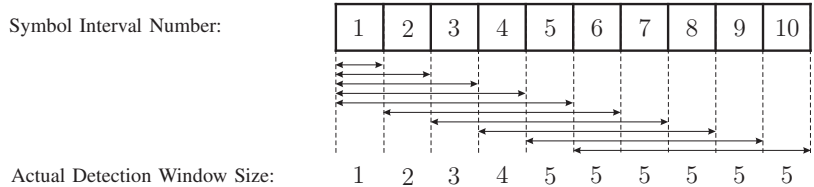}{
\psfrag{k}[l][c][0.8]{Symbol Interval Number:}
\psfrag{K}[l][c][0.8]{Actual Detection Window Size:}
\psfrag{1}[c][c][1]{$1$}
\psfrag{2}[c][c][1]{$2$}
\psfrag{3}[c][c][1]{$3$}
\psfrag{4}[c][c][1]{$4$}
\psfrag{5}[c][c][1]{$5$}
\psfrag{6}[c][c][1]{$6$}
\psfrag{7}[c][c][1]{$7$}
\psfrag{8}[c][c][1]{$8$}
\psfrag{9}[c][c][1]{$9$}
\psfrag{10}[c][c][1]{$10$}
} } \vspace{-0.1cm}
\caption{Schematic illustration of the different detection window sizes required for the DF detection for $K=5$ and $B=10$. In fact, the DF detector employs the desired detection window size of $K=5$  only for the last $B-K+1=6$ symbol intervals in each block of $B$ symbols where the CSI in assumed to be constant. \vspace{-0.3cm}}
\label{Fig:Block}
\end{figure}

\textit{DF Window Size:} In order to employ the proposed DF detector with window size $K$, knowledge of the $K-1$ previously detected symbols is required. However, at the beginning of each transmission interval, i.e., for symbol intervals $k<K$, such knowledge is not available. Therefore, for the results presented in this section, we assume that the DF detector detects the first symbol without knowledge of the previous symbols, i.e., with window size one. Then, it detects the second symbol based on the knowledge of the detected symbol in the first symbol interval, i.e., with window size two. We continue this process until symbol interval $K$ where, for all the following symbols $k\geq K$, the previous $K-1$  detected symbols are available and a fixed window size of $K$ can be used for DF detection. Because of the overlapping detection windows, we implicitly have to assume that the CSI is constant for $B\geq K$ symbol intervals such that a DF detector with window size $K$ can make $B-K+1$ symbol decisions. For the results shown in this section, we consider the BER for only the $B-K+1$ symbol intervals for which the DF detector actually employs a detection window of size $K$. Fig.~\ref{Fig:Block} schematically shows the different detection window sizes used for  DF detection assuming $K=5$ and $B=10$.

\textit{Benchmark Scheme:} We note that the non-coherent detector in \cite{NonCoherent_MC} was designed assuming an additive white Gaussian noise (AWGN) channel model. As discussed in Subsection~II-A, the Poisson distribution for the observed number of molecules at the receiver is a more accurate model for the diffusive MC channel than the Gaussian distribution \cite{HamidJSAC}.  Additionally, the signal model assumed in \cite{NonCoherent_MC} is different from that considered in this paper. Therefore, we have not included the detection scheme in \cite{NonCoherent_MC} as a benchmark in this section as a direct comparison would not~be~fair. Instead we use the optimal coherent detector, cf. Subsection~III.A, as a benchmark scheme which in fact constitutes a performance upper bound for any non-coherent detector.

\subsection{Performance Evaluation}

\begin{figure*}[!tbp]
  \centering  
  \begin{minipage}[b]{0.47\textwidth}
  \centering
\hspace{-0.8cm}
\resizebox{1.1\linewidth}{!}{\psfragfig{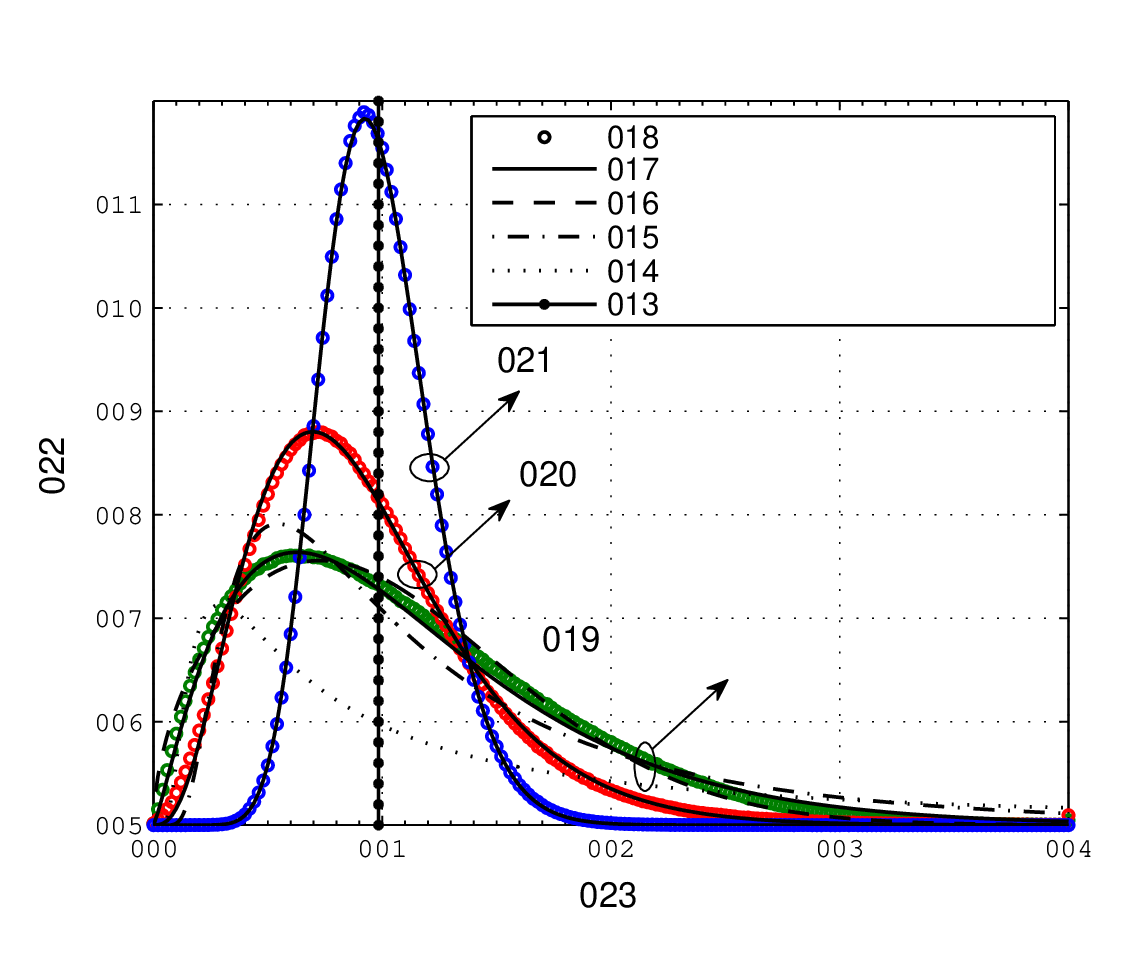}} \vspace{-0.9cm}
\caption{The histogram of $\bar{c}_{\mathtt{s}}$ (exact PDF) and the Gamma PDF for  Scenarios~1-3. Additionally,  the Nakagami, inverse Gaussian, and Levy distributions are plotted only for Scenario~3 to keep the figure readable. \vspace{-0.3cm}}
\label{Fig:GammaPDF}
  \end{minipage}
    \hfill
  \begin{minipage}[b]{0.1\textwidth}
  \end{minipage}
  \hfill
  \begin{minipage}[b]{0.47\textwidth}
  \centering
  \hspace{-0.8cm}
\resizebox{1.1\linewidth}{!}{\psfragfig{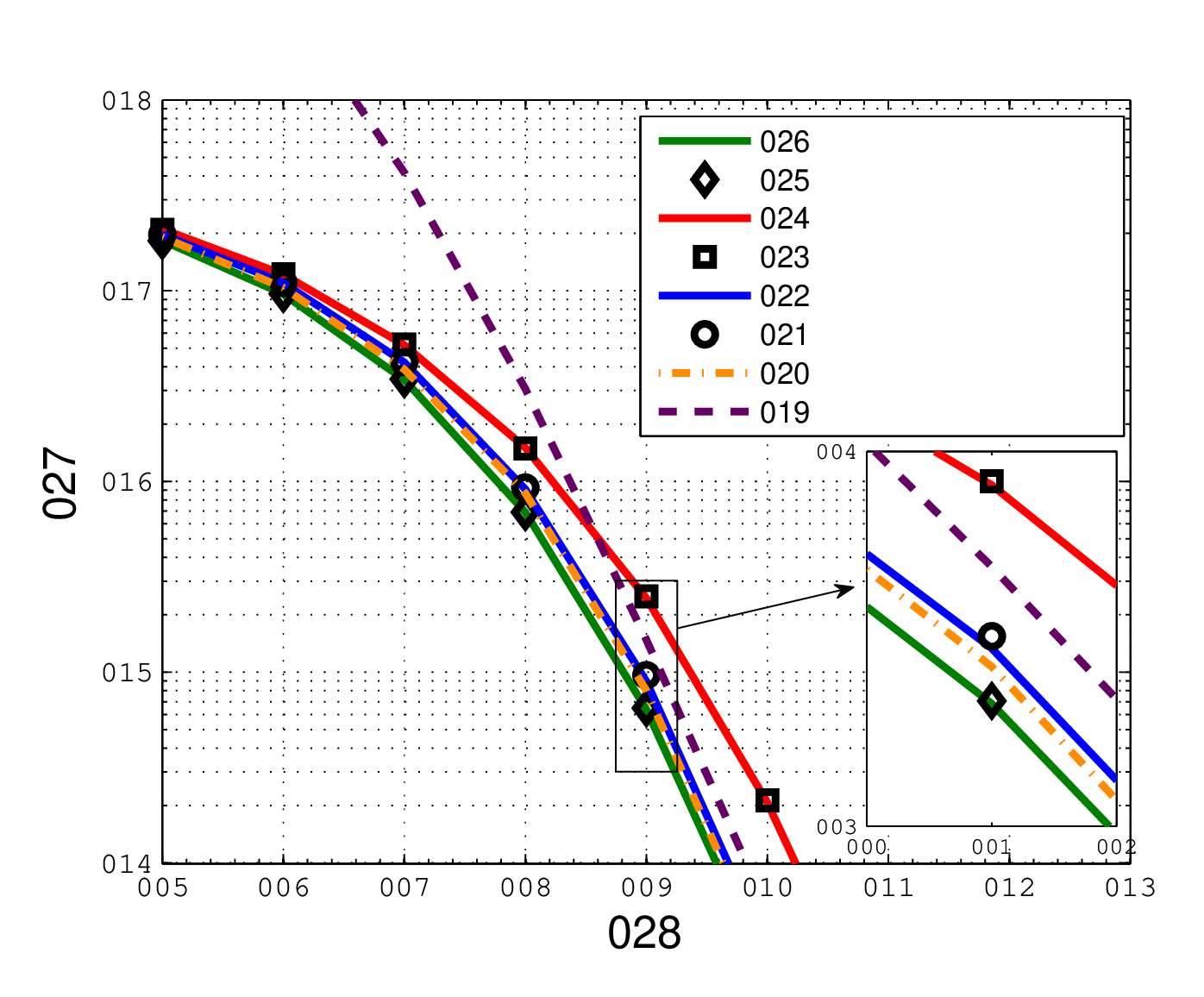}} \vspace{-0.9cm}
\caption{Bit error rate versus the SNR in dB for Scenario~1 and $K=10$.\vspace{-0.3cm} }
\label{Fig:BER_Bound}
  \end{minipage}
    \hfill
  \begin{minipage}[b]{0.02\textwidth}
  \end{minipage}\vspace{-0.4cm}
\end{figure*}

\begin{figure*}[!tbp]
  \centering
  \begin{minipage}[b]{0.47\textwidth}
  \centering
\hspace{-0.8cm}
\resizebox{1.14\linewidth}{!}{\psfragfig{}} \vspace{-0.8cm}
\caption{Bit error rate versus $\kappa$ for Scenarios~1-3, $K=5$, and $\mathtt{SNR}=10$~dB. All curves were obtained via simulation. \vspace{-0.3cm}}
\label{Fig:BER_Kappa}
  \end{minipage}
    \hfill
  \begin{minipage}[b]{0.1\textwidth}
  \end{minipage}
  \hfill
  \begin{minipage}[b]{0.47\textwidth}
  \centering
  \hspace{-0.8cm}
\resizebox{1.1\linewidth}{!}{\psfragfig{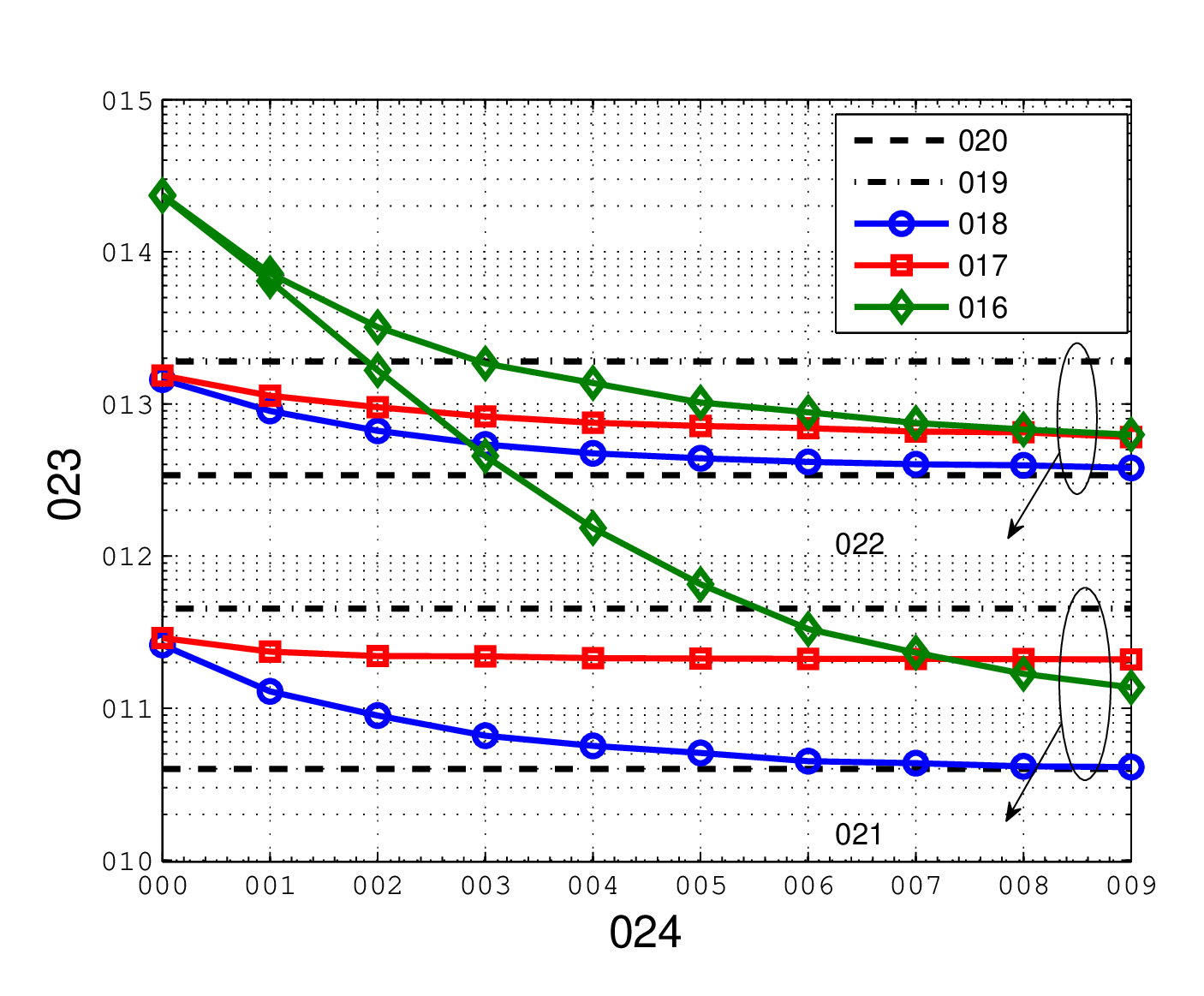}} \vspace{-0.9cm}
\caption{Bit error rate versus the size of the detection window for Scenarios~2 and $\mathtt{SNR}=\{10,20\}$~dB. All curves were obtained via simulation. \vspace{-0.3cm}}
\label{Fig:BER_K}
  \end{minipage}
    \hfill
  \begin{minipage}[b]{0.02\textwidth}
  \end{minipage}\vspace{-0.4cm}
\end{figure*}

\begin{figure*}[!tbp]
  \centering
  \begin{minipage}[b]{0.47\textwidth}
  \centering
\resizebox{1.1\linewidth}{!}{\psfragfig{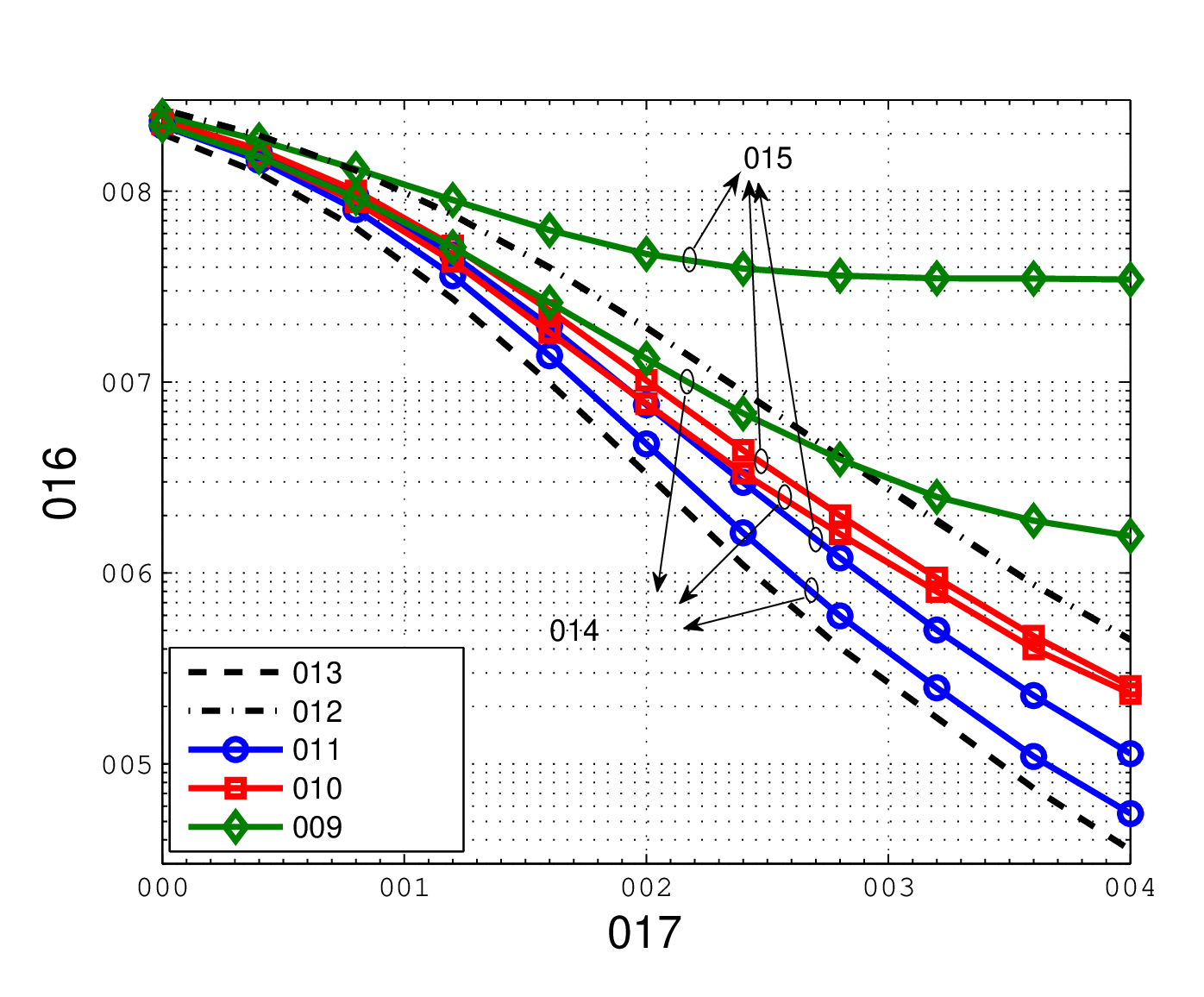}} \vspace{-0.9cm}
\caption{Bit error rate versus the SNR in dB for Scenario~2 and $K\in\{5,10\}$. All curves were obtained via simulation. \vspace{-0.3cm} }
\label{Fig:BER_SNR}
  \end{minipage}
    \hfill
  \begin{minipage}[b]{0.1\textwidth}
  \end{minipage}
  \hfill
  \begin{minipage}[b]{0.47\textwidth}
  \centering
\resizebox{1.1\linewidth}{!}{\psfragfig{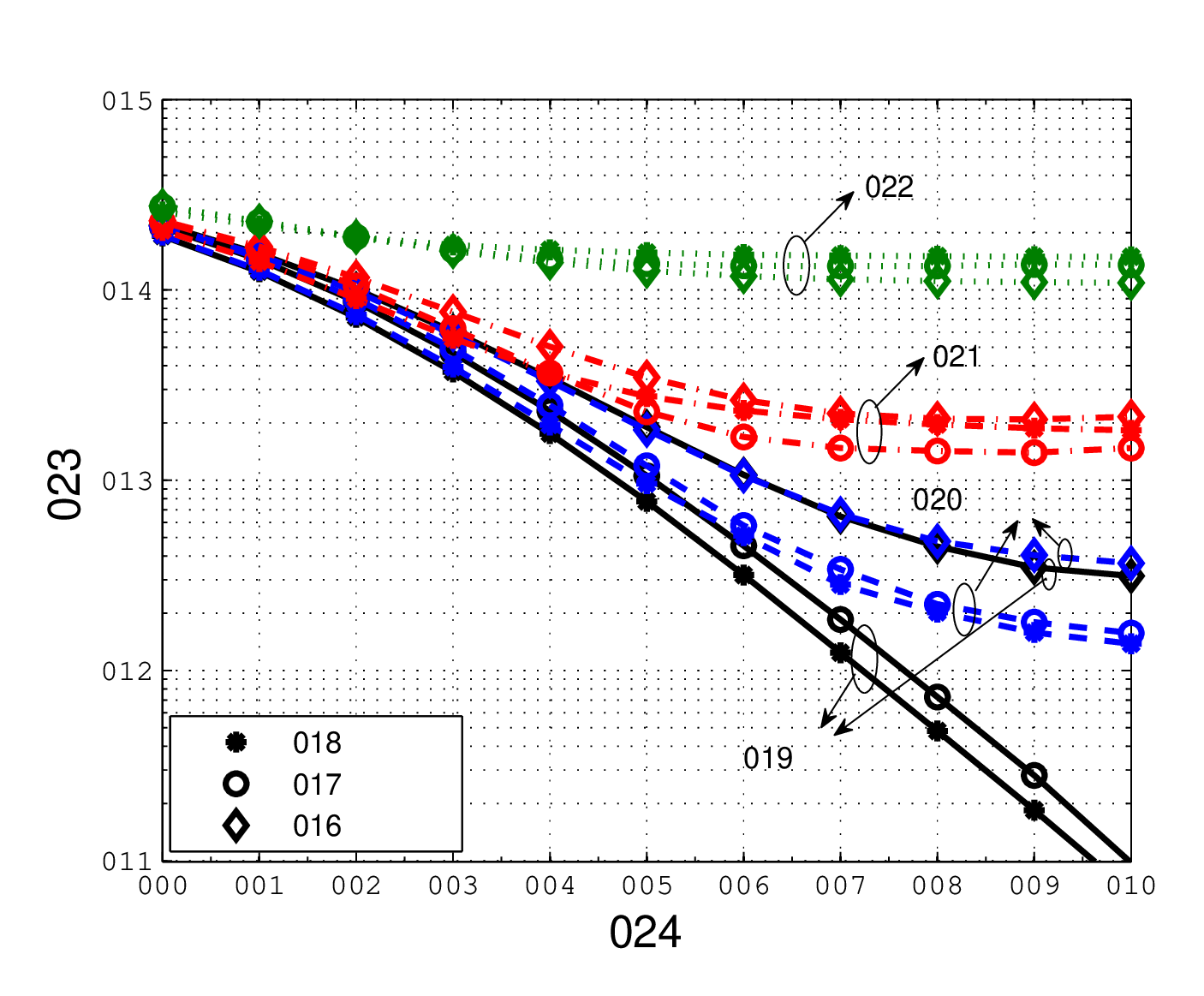}} \vspace{-0.9cm}
\caption{Bit error rate versus the $\mathtt{SNR}^{\mathtt{ext}}$ in dB for Scenario~3 and $K=10$. All curves were obtained via simulation.\vspace{-0.3cm} }
\label{Fig:BER_ISI}
  \end{minipage}
    \hfill
  \begin{minipage}[b]{0.02\textwidth}
  \end{minipage}\vspace{-0.4cm}
\end{figure*}

In this subsection, we first verify the accuracy of the proposed Gamma distribution, cf. Subsection~III-B, in Fig.~\ref{Fig:GammaPDF}, and our analytical derivations, cf. Section~IV, in Fig.~\ref{Fig:BER_Bound}. Subsequently,  in Figs.~\ref{Fig:BER_Kappa}-\ref{Fig:BER_ISI}, we evaluate the performance of the proposed non-coherent detectors for different system parameters.

Fig.~\ref{Fig:GammaPDF} shows the histogram of the CSI, $\bar{c}_{\mathtt{s}}$, obtained by Monte Carlo simulation, and the corresponding Gamma PDF approximation for the three considered stochastic scenarios. Additionally, the result for the case when all the underlying channel parameters in (\ref{Eq:Consentration}) assume their nominal values given in Table~I, i.e., when the channel is deterministic, are shown. The optimal parameters of the Gamma distribution are also shown in Fig.~\ref{Fig:GammaPDF}  and found using the search procedure presented in Subsection~III-A and Remark~\ref{Remk:Search} with $w(x)=1,\,\,\forall x$, and $\delta=0.5$. For Scenario~3,  Nakagami, inverse Gaussian, and Levy distribution approximations are also plotted with their parameters optimized based on a similar search procedure as proposed in Subsection~III-A for the Gamma PDF. We visually observe a very close match between the histogram (exact PDF) and the Gamma PDF approximation for all three scenarios. Moreover, Fig.~\ref{Fig:GammaPDF} suggests that for Scenario~3, the Gamma distribution is a better match to the exact distribution than the Nakagami, inverse Gaussian, and Levy distributions. Note that although the  Nakagami distribution seems to also accurately match the exact distribution in Fig.~\ref{Fig:GammaPDF}, it does not lead to a closed-form expression for the ML detection metric, and hence, at least  for the purpose of this paper, the Nakagami distribution is not a good option. We note that the variances of the channel parameters $(\sigma_{D},\sigma_{v_{\parallel}},\sigma_{v_{\perp}},\sigma_{\bar{c}_{\mathtt{e}}})$ are larger for Scenario~3 compared to Scenario~1, i.e., the underlying channel parameters for Scenario~3 are more random compared to Scenario~1.   From Fig.~\ref{Fig:GammaPDF}, we observe that as the randomness in the MC channel increases, i.e., from Scenario~1 to Scenario~3, the mean of the CSI decreases and its variance increases.

In Fig.~\ref{Fig:BER_Bound}, we verify the simulation results using the performance analysis developed in Section~IV. In particular, we show the BER versus the SNR in dB for Scenario~1 and $K=10$. For the symbol-by-symbol detectors, i.e., the optimal coherent ML (C-ML) detector and the optimal non-coherent SS (NC-SS) detector, we observe that the analytical results obtained from (\ref{Eq:BER_MLSS}) match perfectly with the simulation results. The accuracy of the proposed Gamma distribution to model the considered stochastic MC channel was verified in Fig.~\ref{Fig:GammaPDF} by showing the true and approximated PDFs. In Fig.~\ref{Fig:BER_Bound}, we verify the Gamma approximation in terms of the resulting BER performance. In particular, we show results where the optimal MS detection metric is obtained via expectation over $10^6$ realization of the true CSI, i.e., Monte Carlo simulation, and analytically using the proposed Gamma distribution, cf. (\ref{Eq:Metric_Gamma}). From Fig.~\ref{Fig:BER_Bound}, we observe  that the BERs of the non-coherent MS detectors which employ the optimal metric (NC-MS-OM) and approximated metric (NC-MS-AM)  are almost identical. Furthermore,  we show the union bound (UB) in (\ref{Eq:PEP_r}) and the BER of the genie-aided DF (GA-DF) detector in (\ref{Eq:BER_MLDF}) as an upper bound and a lower bound for the BER of the optimal MS detector, respectively. Note that the union bound becomes a tight upper bound for large SNRs whereas the BER of the genie-aided DF detector is a tight lower bound for all SNR values considered in Fig.~\ref{Fig:BER_Bound}.

In Fig.~\ref{Fig:BER_Kappa}, we show the BER versus $\kappa=\frac{B}{K}$, see Subsection~V.B, for the three considered scenarios, $K=5$, and $\mathtt{SNR}=10$ dB. Note that the actual CSI in (\ref{Eq:Consentration}) is used for simulation whereas the approximated Gamma distribution is employed for calculation of the detection metrics for the proposed MS and DF detectors by using (\ref{Eq:Metric_Gamma}), where the corresponding curves in Fig.~\ref{Fig:BER_Kappa} are denoted by NC-MS-AM and NC-DF-AM, respectively. We observe that the optimal MS detector with detection window size $B$ outperforms the DF detector for all $\kappa \geq 1$ and approaches the performance of the coherent ML detector as $\kappa\to\infty$. For small $\kappa$, the BER of the DF detector is higher than the BER of the optimal MS detector with the same detection window size $K$, however, as $\kappa$ increases, the DF detector outperforms the optimal MS detector with detection window size $K$ by a small margin. This is due to the fact that the DF detector implicitly exploits the property that the CSI is fixed for $B\geq K$ symbol intervals whereas the decision of the optimal MS detector is independent for each detection window size $K$.   Moreover, from Fig.~\ref{Fig:BER_Kappa}, we observe that the BER increases for all considered detectors as the MC channel becomes more stochastic, i.e., from Scenario~1 to Scenario~2 to Scenario~3.

In Fig.~\ref{Fig:BER_K}, we show the BER versus the detection window size $K$ for Scenario~2 and $\mathtt{SNR}\in\{10,20\}$ dB.  We assume $\kappa=1$ for the DF detector. Note that  the BERs of  the proposed detectors decrease and finally converge to the  lower bound provided by the coherent ML detector as $K\to\infty$. Furthermore, the gap between the BER of the optimal non-coherent MS detector and the coherent ML detector is small for $K=20$, which reveals the effectiveness of the optimal  MS detector, although no resources are spent for training  and CSI acquisition.  Moreover, the gap between the BER of the proposed optimal MS detector and the proposed suboptimal non-coherent blind (NC-BL) detector decreases for  larger values of $K$, which confirms the effectiveness of the proposed suboptimal blind detector for large detection window sizes. Furthermore, as expected, the performance of all detection schemes in Fig.~\ref{Fig:BER_K} is better for $\mathtt{SNR}=20$ dB compared to $\mathtt{SNR}=10$ dB.
 
 In Fig.~\ref{Fig:BER_SNR}, we plot the BER versus the SNR in dB for Scenario~2 and $K\in\{5,10\}$. In this figure, we observe that as the SNR increases, the BER improves for all considered detectors. We note that as $\mathtt{SNR}\to\infty$, the BER of the proposed blind detector saturates to a certain error floor. This is due to the fact that for the \textit{blind} CSI estimator in (\ref{Eq:CIR_Est}), we assume that the percentages of ones and zeros in a given detection window are exactly $50\%$, which is not always true, especially for small values of $K$. This introduces an inherent CSI estimation error and leads to the aforementioned error floor for the proposed blind detector. In contrast, none of the other detectors has a BER error floor. Note that as $K$ increases, the BER of the optimal non-coherent MS detector decreases and approaches the BER of the coherent ML detector.  We can also observe from Fig.~\ref{Fig:BER_SNR} that the optimal MS detector outperforms the proposed suboptimal blind detector, particularly for small $K$, but the gap between the BERs of these two detectors decreases as $K$ increases.

Recall that we assume ISI-free transmission in the  system model adopted in this paper. However, although the ISI may be considerably reduced using, e.g., the methods in \cite{Adam_Enzyme} and   \cite{Nariman_Acid}, some residual ISI always exists as the length of the symbol intervals is finite. Therefore, in the system model, we assumed that the effect of the residual ISI is included in $c_{\mathtt{n}}[k]$ and is sufficiently small compared to the other components in $c_{\mathtt{n}}[k]$ such that $c_{\mathtt{n}}[k]$ is (approximately) independent of the signal component $c_{\mathtt{s}}[k] $.  In Fig.~\ref{Fig:BER_ISI}  we study the BER performance loss if this assumption does not hold. To this end, we consider a finite symbol interval length, denoted by $T_{\mathtt{symb}}$, and assume that the contribution of the ISI from the previous symbol intervals is present. The detectors treat the ISI as noise, i.e., $\bar{c}_{\mathtt{n}}=\bar{c}_{\mathtt{n}}^{\mathtt{ext}}+\bar{c}_{\mathtt{n}}^{\mathtt{ISI}}$ where $\bar{c}_{\mathtt{n}}^{\mathtt{ext}}$ is the mean of the external noise and $\bar{c}_{\mathtt{n}}^{\mathtt{ISI}}$ is the expected ISI given by
\begin{IEEEeqnarray}{lll} \label{Eq:ISI_Exp}
\bar{c}_{\mathtt{n}}^{\mathtt{ISI}} = \mathbbmss{E}\left\{\sum_{l=2}^{\infty}s[k-l+1]\bar{c}_{\mathtt{s},l}\right\} = \frac{1}{2}\sum_{l=2}^{\infty}\bar{c}_{\mathtt{s},l},
\end{IEEEeqnarray}
where $\bar{c}_{\mathtt{s},l}$ is the $l$-th channel tap, i.e., if the transmitter releases $N^{\mathtt{TX}}$ molecules in symbol interval $k$, $\bar{c}_{\mathtt{s},l}$ is the \textit{expected} number of molecules arriving at the destination in symbol interval $k+l-1$. Therefore, $\bar{c}_{\mathtt{s},1}$ is the channel tap for the desired signal and $\bar{c}_{\mathtt{s},l},\,\,l\geq 2$, are channel taps that create ISI.

 In Fig.~\ref{Fig:BER_ISI}, we show the BER versus $\mathtt{SNR}^{\mathtt{ext}}=\frac{\bar{c}_{\mathtt{s},1}}{\bar{c}_{\mathtt{n}}^{\mathtt{ext}}}$ in dB for Scenario~3, $K=10$, and symbol intervals $T_{\mathtt{symb}}\in\{2,5,10\}\times T_{\mathtt{max}}$ where $T_{\mathtt{max}}=\underset{t>0}{\mathrm{argmax}}\,\,\bar{c}_{\mathtt{s}}(t)$ and $\bar{c}_{\mathtt{s}}(t)$ is obtained from (\ref{Eq:Consentration}) assuming the nominal values of the MC system parameters in Table~I. We observe from Fig.~\ref{Fig:BER_ISI} that as $\mathtt{SNR}^{\mathtt{ext}}\to\infty$, the BERs of all detectors saturate to a BER floor if ISI is present. Moreover, due to the severe mismatch between the system model assumed for development of the considered detectors and the actual simulated MC channel for the strong ISI case, the relative performance of these detectors is not obvious. In fact, we observe from Fig.~\ref{Fig:BER_ISI} that  for the considered system parameters and $T_{\mathtt{symb}}=2 T_{\mathtt{max}}$, the performance of the coherent ML detector considerably deteriorates  compared to that of the MS detector and, similarly, the performance of the MS detector degrades compared to that of the blind detector. This suggests that the coherent ML detector is more sensitive to the considered system model mismatch compared to the MS detector and, similarly, the  MS detector is more sensitive compared to the blind detector. Finally, Fig.~\ref{Fig:BER_ISI} shows that as the ISI becomes weaker, i.e., as $T_{\mathtt{symb}}$ increases, the BERs of all detectors approach the respective BERs for the ISI-free case\footnote{We note that optimal non-coherent detection for the case when strong ISI is present is an important research problem which is beyond the scope of this paper and left for future work.}.

\section{Conclusions}

We have derived the optimal non-coherent  MS and SS detectors as well as a non-coherent DF detector  which do not require instantaneous CSI knowledge. As compared to the coherent detectors previously studied, the proposed non-coherent detectors may be preferable in practical scenarios  since the complexity and the overhead associated with CSI acquisition are avoided.  In order to further reduce the complexity of our detectors, we proposed an approximate detection metric and a low-complexity suboptimal blind detector. We further derived an analytical expression for the BER of the optimal SS detector, and a lower bound and an upper bound  for the BER of the optimal MS detector. Simulation results confirmed the analysis and showed that the proposed optimal MS detector outperforms the suboptimal blind detector, particularly for small detection window sizes. However, as the size of the detection window increases, the performances of the proposed optimal and suboptimal detectors converge to that of the benchmark coherent ML detector which requires perfect CSI. This demonstrates the effectiveness of the proposed detection schemes. 


\appendices

\section{Proof of Theorem~\ref{Theo:MLMS}}\label{App:Theo_MLMS}

Using sets $\mathcal{K}_1$ and $\mathcal{K}_0$, (\ref{Eq:MLMS}) can be simplified to
\begin{align} \label{Eq:MLMS_sets}
  \mathbf{s}^{\mathtt{MS}} =  \underset{\mathbf{s}\in\mathcal{A}}{\mathrm{argmax}} \,\,  \int_{\bar{c}_{\mathtt{s}}\geq 0} \int_{\bar{c}_{\mathtt{n}}\geq 0} 
 & \left( \bar{c}_{\mathtt{s}} + \bar{c}_{\mathtt{n}} \right)^{\sum_{k\in\mathcal{K}_1} r[k]}
  \bar{c}_{\mathtt{n}}^{\sum_{k\in\mathcal{K}_0} r[k]} \nonumber \\
  & \times \exp\left(-|\mathcal{K}_1|\bar{c}_{\mathtt{s}}  - K \bar{c}_{\mathtt{n}} \right)
  f_{(\bar{c}_{\mathtt{s}},\bar{c}_{\mathtt{n}})}(\bar{c}_{\mathtt{s}},\bar{c}_{\mathtt{n}}) \mathrm{d}\bar{c}_{\mathtt{s}}\mathrm{d}\bar{c}_{\mathtt{n}}, \quad
\end{align}
where the term $r[k]!$ in (\ref{Eq:MLMS}) is removed in (\ref{Eq:MLMS_sets}) since it does not affect the MS detection result. Moreover, the optimal MS detector can be written equivalently in  expectation form as
\begin{align} \label{Eq:MLMS_expect}
\vspace{-2cm}  \mathbf{s}^{\mathtt{MS}} = \underset{\mathbf{s}\in\mathcal{A}}{\mathrm{argmax}} \,\, \mathbbmss{E}_{\bar{\mathbf{c}}}\bigg\{ 
  \underset{A(\mathbf{s},\mathbf{r})}{\underbrace{\left( \bar{c}_{\mathtt{s}} + \bar{c}_{\mathtt{n}} \right)^{\sum_{k\in\mathcal{K}_1} r[k]}
  \bar{c}_{\mathtt{n}}^{\sum_{k\in\mathcal{K}_0} r[k]}}}\, \underset{B(\mathbf{s},\mathbf{r})}{\underbrace{\exp\left(-|\mathcal{K}_1|\bar{c}_{\mathtt{s}}  - K \bar{c}_{\mathtt{n}} \right)}} \bigg\}. \quad
\end{align}
We can conclude the following properties from the MS detection metric $\Lambda^{\mathtt{MS}}(\mathbf{s},\mathbf{r})=\mathbbmss{E}_{(\bar{c}_{\mathtt{s}}, \bar{c}_{\mathtt{n}})}\{A(\mathbf{s},\mathbf{r}) $ $ B(\mathbf{s},\mathbf{r})\}$. First, with respect to the data sequence $\mathbf{s}$, term $B(\mathbf{s},\mathbf{r})$ is only a function of the number of ones in $\mathbf{s}$. Second, for a given number of ones in $\mathbf{s}$, term $A(\mathbf{s},\mathbf{r})$ is maximized if the ones in $\mathbf{s}$ correspond to the largest  elements of the observation vector $\mathbf{r}$. Note that these two properties hold for any given CSI $\bar{\mathbf{c}}$. Hence, they also hold after the expectation operation with respect to $\bar{\mathbf{c}}$, i.e., $\mathbbmss{E}_{\bar{\mathbf{c}}}\{A(\mathbf{s},\mathbf{r}) B(\mathbf{s},\mathbf{r})\}$. Therefore, we can avoid searching over all $\mathbf{s}\in\mathcal{A}$, and instead, find the optimal threshold $n_1 \in\{0,1,\dots,K\}$ which sets the elements of $\mathbf{s}$ corresponding to the $n_1$ largest elements of $\mathbf{r}$ to one and the remaining elements  to zero. Moreover, we can further simplify the MS detection metric in (\ref{Eq:MLMS_expect}) as 
\begin{align} \label{Eq:MLMS_bino}
\displaystyle\Lambda^{\mathtt{MS}}(\mathbf{s},\mathbf{r})
 &\displaystyle\overset{(a)}{=}  \mathbbmss{E}_{(\bar{c}_{\mathtt{s}},\bar{c}_{\mathtt{n}})}\left\{
  \left[\sum_{i=0}^{N_1} {N_1 \choose i} \bar{c}_{\mathtt{s}}^{N_1-i} \bar{c}_{\mathtt{n}}^i \right]  \bar{c}_{\mathtt{n}}^{N_0} 
  \exp\left(- \big|\mathcal{K}_1\big|\bar{c}_{\mathtt{s}}  - K \bar{c}_{\mathtt{n}} \right) \right\}  \quad \nonumber \\
  & \displaystyle\overset{(b)}{=} \sum_{i=0}^{N_1} {N_1 \choose i} \mathbbmss{E}_{\bar{c}_{\mathtt{s}}}\left\{
   \bar{c}_{\mathtt{s}}^{N_1-i} e^{- n_1 \bar{c}_{\mathtt{s}}} \right\} 
  \mathbbmss{E}_{\bar{c}_{\mathtt{n}}}\left\{   \bar{c}_{\mathtt{n}}^{N_0+i} e^{- K \bar{c}_{\mathtt{n}}}   \right\} , \hspace{-0.1cm}
\end{align}
where in equality $(a)$, we employ the Binomial expansion, i.e., $(x+y)^n = \sum\limits_{i = 0}^n{ {n \choose i} x^{n-i} y^i }$ with ${n \choose i}=\frac{n!}{i!(n-i)!}$,  and in  equality $(b)$,  we use $n_1=|\mathcal{K}_1|$ and the assumption that  RVs $\bar{c}_{\mathtt{s}}$ and $\bar{c}_{\mathtt{n}}$ are independent. Furthermore, recalling that  $\Pr\{s[k]=1\}=\Pr\{s[k]=0\}=0.5$ holds, the optimal threshold, denoted by $\zeta^{\mathtt{MS}}(\mathbf{r})$, approaches $K/2$ as $K\to\infty$. This concludes the proof.

\section{Proof of Corollary~\ref{Corol:MLSS}} \label{App:Corol_MLSS}

For the special case of the symbol-by-symbol detection, we obtain two detection metrics from (\ref{Eq:MLMS_expect}) corresponding to $s[k]=1$ and $s[k]=0$, respectively, i.e.,  
\begin{IEEEeqnarray}{lll} \label{Eq:MLSS_expect}
  \Lambda(s[k]) = \begin{cases} \mathbbmss{E}_{(\bar{c}_{\mathtt{s}},\bar{c}_{\mathtt{n}})}\left\{ \left( \bar{c}_{\mathtt{s}} + \bar{c}_{\mathtt{n}} \right)^{r[k]} e^{- (\bar{c}_{\mathtt{s}} + \bar{c}_{\mathtt{n}} ) } \right\},\quad &\mathrm{if}\,\, s[k] = 1 \\
  \mathbbmss{E}_{\bar{c}_{\mathtt{n}}}\left\{ \bar{c}_{\mathtt{n}}^{r[k]} e^{ - \bar{c}_{\mathtt{n}} } \right\},\quad &\mathrm{otherwise}
  \end{cases}
\end{IEEEeqnarray}
Now, using Lemma~\ref{Lem:func}, we first show that $\frac{\Lambda(s[k]=1)}{\Lambda(s[k]=0)}$ is a monotonically increasing function of $r[k]$. Moreover, for $r[k]=0$ and $r[k]\to\infty$, we obtain that $\Lambda(s[k]=1)<\Lambda(s[k]=0)$ and $\Lambda(s[k]=1)>\Lambda(s[k]=0)$ hold, respectively. Therefore, there exists a unique threshold for $r[k]$ below which $\Lambda(s[k]=1)<\Lambda(s[k]=0)$ holds. 

\begin{lem}\label{Lem:func}
If $\frac{f_n(x)}{g_m(x)}$ is a monotonically increasing function of $x$ for all possible pairs of $(m,n)$, then function $\frac{\sum_{n}f_n(x)}{\sum_m g_m(x)}$ is also monotonically increasing in $x$.
\end{lem}
\begin{IEEEproof}
Please refer to Appendix~\ref{App:Lem_func}.
\end{IEEEproof}
In order to apply the result of Lemma~\ref{Lem:func} to $\frac{\Lambda(s[k]=1)}{\Lambda(s[k]=0)}$, we first note that the expectation terms $\mathbbmss{E}_{x}\{f(x)\}$ can be written in summation form as $\sum_n \Pr\{x_n\} f(x_n) $ by discretizing the domain of $x$ into a set $\{x_1,x_2,\dots\}$.  Therefore, we have $\Lambda(s[k]=1)\triangleq \sum_n f_n(r[k]) = \sum_n \Pr\{\bar{c}_{\mathtt{s}}\} \Pr\{\bar{c}_{\mathtt{n}}\} \\ \left( \bar{c}_{\mathtt{s}} + \bar{c}_{\mathtt{n}} \right)^{r[k]} $ $e^{- (\bar{c}_{\mathtt{s}} + \bar{c}_{\mathtt{n}} ) }$ and $\Lambda(s[k]=0)\triangleq \sum_m g_m(r[k]) = \sum_n \Pr\{\bar{c}_{\mathtt{n}}\}  \bar{c}_{\mathtt{n}}^{r[k]} e^{-  \bar{c}_{\mathtt{n}} }$. Now, we have to show that $\frac{f_n(r[k])}{g_m(r[k])} = \frac{\Pr\{\bar{c}_{\mathtt{s}}\} \Pr\{\bar{c}_{\mathtt{n}}\} \left( \bar{c}_{\mathtt{s}} + \bar{c}_{\mathtt{n}} \right)^{r[k]} e^{- (\bar{c}_{\mathtt{s}} + \bar{c}_{\mathtt{n}} ) }}{\Pr\{\bar{c}_{\mathtt{n}}\}  \bar{c}_{\mathtt{n}}^{r[k]} e^{-  \bar{c}_{\mathtt{n}} }} = \Pr\{\bar{c}_{\mathtt{s}}\} \left(1+\frac{\bar{c}_{\mathtt{s}}}{\bar{c}_{\mathtt{n}}}\right)^{r[k]} e^{-\bar{c}_{\mathtt{s}}}$ is a monotonically increasing function of $r[k]$ for all $\bar{c}_{\mathtt{s}}$ and $\bar{c}_{\mathtt{n}}$ which straightforwardly holds.

To further simplify the SS detection metric under hypothesis $s[k]=1$, we employ the Binomial expansion which leads to 
\begin{IEEEeqnarray}{lll} \label{Eq:MLSS_expect_1}
  \Lambda(s[k]=1) = \sum\limits_{i = 0}^{r[k]}  {r[k] \choose i} 
  \mathbbmss{E}_{\bar{c}_{\mathtt{s}}}\left\{ \bar{c}_{\mathtt{s}}^{r[k]-i}e^{-\bar{c}_{\mathtt{s}}} \right\}
   \mathbbmss{E}_{\bar{c}_{\mathtt{n}}}\left\{ \bar{c}_{\mathtt{n}}^{i}e^{-\bar{c}_{\mathtt{n}}} \right\}.
\end{IEEEeqnarray}
The above results are concisely  stated in Corollary~\ref{Corol:MLSS} which concludes the proof.

\section{Proof of Lemma~\ref{Lem:func}} \label{App:Lem_func}

If $\frac{f_n(x)}{g_m(x)}$ is a monotonically increasing function of $x$, its derivative with respect to $x$ has to be positive, i.e.,
\begin{IEEEeqnarray}{lll}\label{Eq:func_ind} 
\frac{\partial}{\partial x} \left(\frac{f_n(x)}{g_m(x)}\right) = \frac{f'_n(x)g_m(x)-f_n(x)g'_m(x)}{(g_m(x))^2} > 0,\quad \forall n,m.
\end{IEEEeqnarray}
In other words, $f'_n(x)g_m(x)-f_n(x)g'_m(x)> 0$ has to hold for $\forall n,m$. Using this result, the sufficient condition for $\frac{\sum_{n}f_n(x)}{\sum_m g_m(x)}$ to be a monotonically increasing function of $x$ can be shown as
\begin{IEEEeqnarray}{lll} \label{Eq:func_tot} 
\frac{\partial}{\partial x} \left(\frac{\sum_n f_n(x)}{\sum_m g_m(x)}\right) & = \frac{\sum_n f'_n(x)\times \sum_m g_m(x) - \sum_n f_n(x)\times \sum_m g'_m(x) }{(\sum_m g_m(x))^2} \nonumber \\
& \overset{(a)}{=} \frac{\sum_n\sum_m f'_n(x) g_m(x) - \sum_n \sum_m f_n(x)   g'_m(x) }{(\sum_m g_m(x))^2} \nonumber \\
& = \frac{\sum_n\sum_m [f'_n(x) g_m(x) -  f_n(x)   g'_m(x)] }{(\sum_m g_m(x))^2} \overset{(b)}{>} 0,
\end{IEEEeqnarray}
where equality $(a)$ follows the sum-product rule, i.e., $\sum_n x_n \sum_m y_m = \sum_n\sum_m x_n y_m$, and inequality $(b)$ follows from (\ref{Eq:func_ind}). This completes the proof.

\bibliographystyle{IEEEtran}
\bibliography{Ref_27_07_2017}

\end{document}